\documentclass[reprint,amsmath,amssymb,aps,prx,floatfix]{revtex4-2}
\pdfoutput=1

\usepackage{silence}
\usepackage[colorlinks=true,linkcolor=blue,citecolor=blue,urlcolor=blue]{hyperref}
\usepackage{graphicx}
\usepackage{comment}
\usepackage{epsfig}
\usepackage{epstopdf}
\usepackage[usenames,dvipsnames]{xcolor}
\usepackage{braket}
\usepackage{amsthm}
\usepackage{mathtools}
\usepackage{enumitem}
\usepackage{dsfont}
\usepackage{bbold}
\usepackage{bm}
\usepackage{eurosym}
\usepackage{tabularx}
\usepackage{mdframed}
\usepackage{cases}
\usepackage{adjustbox}
\usepackage{diagbox}
\usepackage{ragged2e}
\usepackage[symbol]{footmisc}
\usepackage{afterpage}

\usepackage[most]{tcolorbox}
\newtcolorbox{mybox}[2][]{
               = {yshift=-8pt},
  colback      = blue!6!white,
  colframe     = blue!1!black,
  halign       = flush left,
  fonttitle    = \bfseries\sffamily,
  colbacktitle = blue!90!black,
  title        = #2,#1,
  enhanced,
}

\newcommand{\be}{\begin{equation}}
\newcommand{\ee}{\end{equation}}
\newcommand{\ba}{\begin{aligned}}
\newcommand{\ea}{\end{aligned}}

\newcommand{\bc}{\begin{center}}
\newcommand{\ec}{\end{center}}
\newcommand{\beq}{\begin{equation}}
\newcommand{\eeq}{\end{equation}}
\newcommand{\beqq}{\begin{equation*}}
\newcommand{\eeqq}{\end{equation*}}
\newcommand{\beqa}{\begin{align}}
\newcommand{\eeqa}{\end{align}}
\newcommand{\barr}{\begin{array}}
\newcommand{\earr}{\end{array}}
\newcommand{\bi}{\begin{itemize}}
\newcommand{\ei}{\end{itemize}}

\newcommand{\Tr}{\ensuremath{\,\mathrm{Tr}}}

\newtheorem*{theo*}{Theorem}

\newtheorem*{prot*}{Protocol}

\begin{document}


\title{Rigorous characterization of continuous-variable quantum states via optical parametric amplifiers}

\author{Manthan Badbaria\textsuperscript{1}}
\author{Fumiya Hanamura\textsuperscript{2}}

\author{Maxime Garnier\textsuperscript{3}}

\author{Ulysse Chabaud\textsuperscript{3}}
\email{ulysse.chabaud@inria.fr}
\author{Rajveer Nehra\textsuperscript{1,4,5,6}}
\email{rajveernehra@umass.edu}

\affiliation{$^1$Department of Physics, University of Massachusetts Amherst, Amherst, Massachusetts 01003, USA}
\affiliation{$^2$Department of Applied Physics, School of Engineering, The University of Tokyo, 7-3-1 Hongo, Bunkyo-ku, Tokyo 113-8656, Japan}
\affiliation{$^3$DIENS, \'Ecole Normale Sup\'erieure, PSL University, CNRS, INRIA, 45 rue d'Ulm, Paris 75005, France}

\affiliation{$^4$Department of Electrical and Computer Engineering, University of Massachusetts Amherst, Amherst, Massachusetts 01003, USA}

\affiliation{$^5$College of Information and Computer Science, University of Massachusetts Amherst,
Amherst, Massachusetts 01003, USA}

\affiliation{$^{6}$Materials Science and Engineering Program, University of Massachusetts Amherst, Amherst, MA 01003, USA}

\begin{abstract}
Characterizing non-Gaussian quantum states is of paramount importance for continuous-variable quantum information processing, yet conventional homodyne-measurement-based state tomography remains limited by optical loss, detector efficiency, and measurement bandwidth. Here, we introduce an integrated framework for loss-tolerant characterization and certification using high-gain phase-sensitive optical parametric amplification and power measurements. Our computationally efficient semidefinite programming approach enables faithful reconstruction of parity-symmetric quantum states from amplified quadrature measurements while substantially relaxing detector-efficiency requirements and increasing the measurement bandwidth. We further develop a certification framework that directly quantifies non-Gaussianity via stellar-rank witnesses and Wigner negativity, using the same quadrature-power measurements. We demonstrate the efficacy of the proposed framework through both simulated and experimental data for representative quantum states, including single-photon, Schr\"odinger cat, and Gottesman--Kitaev--Preskill (GKP) states. By unifying loss-tolerant measurements, state tomography, and nonclassical-state certification within a single experimentally accessible framework, our approach provides a practical pathway toward verifying increasingly complex states and can be readily implemented with current quantum photonic technologies.
\end{abstract}

\maketitle


\section{Introduction}
Quantum information processing (QIP) ushers in a new paradigm of next-generation information processors, offering significant advantages over classical counterparts for some specific tasks. Over the last two decades, many physical platforms, including superconducting circuits, trapped ions, quantum dots, nuclear spins, neutral atoms, and photonics, have been investigated for QIP~\cite{clarke2008superconducting,o2009photonic,cirac1995quantum,imamog1999quantum, saffman2010quantum,vandersypen2001experimental}. Photonic QIP provides critical advantages such as room temperature operations, ultrafast clock speeds, compatibility with telecommunication systems, and the ability to easily network multiple modular quantum processors for building large-scale quantum processors~\cite{ladd2010quantum,o2009photonic}.

In particular, continuous-variable (CV) QIP, which encodes quantum information in continuous amplitude and phase
quadratures of an electromagnetic field, provides unprecedented scalability, as exemplified by 
the deterministic generation of large-scale CV cluster states comprising tens of thousands of modes~\cite{Asavanant2019_alt,Larsen2019_alt}. In CVQIP, quantum states are separated into two categories, Gaussian and non-Gaussian~\cite{weedbrook2012gaussian}, based on the shape of their Wigner function---a well-known quasiprobability representation in phase space~\cite{Wigner1932}. Non-Gaussian states are essential to various CVQIP tasks~\cite{eisert2002distilling,Bartlett2002}, including quantum error correction~\cite{fiuravsek2002gaussian}.

An essential subset of non-Gaussian states is those with the Wigner function taking negative values. These states are critical to any quantum computational advantage~\cite{Mari2012} and correspond precisely to the states that lead to contextuality in the standard model of CV quantum computing~\cite {booth2022contextuality}. Moreover, non-Gaussian states such as binomial states~\cite{Binomial_code}, cat states~\cite{cat_code}, and Gottesman--Kitaev-Preskill (GKP) states~\cite{Gottesman2001}, exhibit phase-space symmetries that enable robust information encoding through redundancy. Hereafter, we will use the term \textit{parity-symmetric} to refer to states that are invariant under the action of the parity operator (up to a global phase), which applies to the aforementioned quantum states.

Given their importance for QIP, efficient generation and characterization of non-Gaussian quantum states have garnered increasing attention over the years~\cite{lvovsky2020production,walschaers2021non, endo2023non,eaton2019non,ourjoumtsev2007generation,yanagimoto2023quantum}. 
The stellar rank formalism has emerged as a useful method to categorize non-Gaussian quantum states based on the number of elementary non-Gaussian operations that their preparation requires, specifically photon additions and subtractions, referred to as the stellar rank~\cite{chabaud2020stellar}. Like Wigner negativity, this measure also assesses the potential computational advantage of CV quantum computation~\cite{chabaud2020classical,chabaud2022resources}.

Conventional tomographic methods for fully characterizing CV quantum systems utilize interferometric schemes such as strong- and weak-field homodyne detection with classical low-noise photodetectors and superconducting photon-number-resolving detectors, respectively~\cite{smithey1993measurement, nehra2019state,thekkadath2020tuning}. Since the first experimental demonstration by Smithey \textit{et al.}~\cite{smithey1993measurement}, strong-field balanced homodyne detection (BHD) has been extensively used for performing quantum state tomography~\cite{ourjoumtsev2007generation,endo2023non}. However, its application in nanophotonics poses significant challenges. In nanophotonics, quantum states are typically measured with off-chip detectors, which suffer from coupling losses and non-unit quantum efficiency of the photodetectors~\cite{arrazola2021quantum,peace2022picosecond,yang2021squeezed}. Additionally, BHD bandwidth is limited by the electronic bandwidth ranging from MHz to a few tens of GHz~\cite{tasker2021silicon,bruynsteen2021integrated}. On the other hand, weak-field homodyne detection employs superconducting detectors for photon-number resolved measurements, introducing further complexities in quantum state tomography protocols~\cite{thekkadath2020tuning,nehra2020generalized}.

In this work, we propose using an all-optical loss-tolerant scheme to explore the tradeoff between incomplete measurements, that is, measurements that are not tomographically complete, and loss-tolerant characterization of quantum states for CVQIP. In particular, we propose to replace the use of BHD, which measures field quadratures of the electromagnetic field, by measuring \textit{the absolute value} of a quadrature, which we call \textit{sign-free quadrature measurement}. Although, in general, this measurement is no longer tomographically complete, we show that it still allows for the characterization of a large class of quantum states instrumental for quantum error correction in CVQIP, namely parity-symmetric states, while offering critical advantages such as tolerance to overall detection losses and overcoming bandwidth limitations in homodyne detection.

Our scheme employs a high-gain phase-sensitive optical parametric amplifier (OPA) to measure the \textit{sign-free quadrature measurement} directly by amplifying the microscopic fields to macroscopic levels, followed by quadrature power detection. 
More recently, such measurement techniques have been used to perform loss-tolerant measurements of squeezed states~\cite{shaked_lifting_2018, nehra2022few,takanashi2020all, kalash2022wigner}, SU(1,1) interferometry and distributed sensing~\cite{frascella2019wide,nehra2026all}, and all-optical feed-forward operations~\cite{yamashima2025all}.
We show how the sign-free quadrature measurements can be used to achieve robust and loss-tolerant certification of quantum optical states. Our contribution is two-fold: \textit{(i)} we present a method for performing comprehensive quantum state tomography for parity-symmetric quantum states using a computationally efficient reconstruction procedure based on semidefinite programming (SDP), and \textit{(ii)} we show how to directly estimate linear functions of the measured state, such as the fidelity with target pure states, with rigorous confidence intervals. We demonstrate the applicability of our results using experimental homodyne data and numerically simulated data by performing loss-tolerant state tomography and non-classical property witnessing.

The rest of the paper is organized as follows. Section~\ref{sec:expscheme} presents the all-optical experimental scheme we use, which utilizes a high-gain phase-sensitive OPA followed by quadrature power detection to achieve sign-free quadrature measurement. In section~\ref{sec:dataprocessing}, we explain how the sign-free quadrature measurement data can achieve robust quantum state certification, including loss-tolerant quantum state tomography and witnesses for non-classical properties such as Wigner negativity. Our results are supported by experimental and numerically simulated data, which we present in the paper. Section~\ref{sec:conclusion} provides the concluding remarks and outlines future directions for this research.


\section{High-gain OPAs for direct quantum state characterization}
\label{sec:expscheme}

\begin{figure}[t]
    \centering

    \includegraphics[width=\columnwidth]{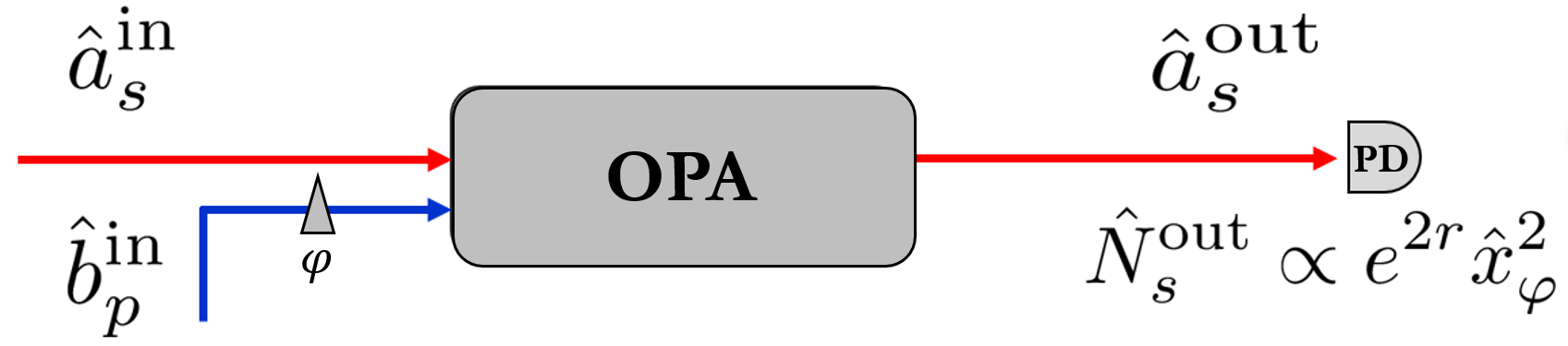}
    \caption{Illustration of the experimental scheme for sign-free quadrature measurement. The operators represent the input signal and pump modes $\hat{a}^{\text{in}}_s$   $\hat{b}^{\text{in}}_p$, respectively. 
    The unknown quantum state is first amplified with a high-gain OPA. Subsequently, a quadrature power measurement is performed on the amplified state, directly detecting the squared quadrature selected by the pump phase.}
    \label{fig:expscheme}
\end{figure}

In this section, we discuss the scheme we use for quantum state certification through sign-free quadrature measurements. The scheme, illustrated in Fig.~\ref{fig:expscheme}, consists of a high-gain phase-sensitive OPA with the signal and pump modes associated with the photon annihilation operators $\hat{a}_s$ and $\hat{b}_p$, respectively. Here we consider the parametric approximation, which assumes that the pump mode can be treated classically and its dynamics remain unchanged throughout the time evolution~\cite{d1999parametric}. In the Heisenberg picture, the evolution of the signal mode is given by the Bogoliubov transformation.
\begin{equation}
    \hat{a} \rightarrow \hat{a}^{}_s\cosh r  + e^{i\phi}\hat{a}^{\dag}_s\sinh r,
    \label{eq:OPA_H}
\end{equation}
where $\varphi$ and $r$ correspond to the amplification phase and OPA gain parameter with gain $G = e^{2r}$, respectively. Likewise, one can calculate the evolved photon-number operator. Here, $\varphi$ represents the amplification phase, and $ r$ is the OPA gain parameter, with the gain $ G$ being the exponential of $ 2r$. We have
\begin{equation}
    \hat{N} = c_1 (\hat{a}^{2}_se^{-i\varphi} + \hat{a}_s^{\dagger2}e^{i\varphi})+ c_2\hat{a}_s^\dagger\hat{a}_s + c_3.
\end{equation}
Here we have set $c_1=\cosh r\sinh r$, $c_2 = \cosh^2r+\sinh^2r$, and $c_3 = \sinh^2r$. We now focus on computing the positive operator-valued measure (POVM) for the detection scheme by adopting the formalism developed in Ref.~\cite{d649advances}. The moment-generating function of the photocurrent is 
\begin{equation}
    \chi(\lambda) = \text{Tr}[\rho e^{i\lambda\hat{I}}],
\end{equation}
Where $\hat{I} \propto \hat{N}$ and $\rho$ denote the output photocurrent and input density operators, respectively. The photocurrent probability distribution is (see Appendix \ref{app:QST}) 
\begin{equation}
    \label{eq:I_distribution}
    P(I) \!=\! \int_{-\infty}^{
    \infty}\! \frac{d\lambda}{2\pi} e^{-i\lambda I}\chi(\lambda) \!=\! \frac{1}{G}\!\!\int \!\frac{dz}{2\pi}e^{-2izx_\varphi^2} \text{Tr}[\rho e^{-2iz\hat{x}_\varphi^2} ]. 
\end{equation}
Here $\hat{x}_\varphi:= (\hat{a}e^{i\varphi/2}+\hat{a}^\dagger e^{-i\varphi/2})/\sqrt{2}$ is the generalised operator for the amplified quadrature. Up to a rescaling of the outcomes, the corresponding POVM element describing the detection process is
\begin{equation}
    \hat\Pi_\varphi=|x_\varphi\rangle\langle x_\varphi|+|-x_\varphi\rangle\langle-x_\varphi|.
    \label{eq:povm}
 \end{equation}
This method is equivalent to measuring the absolute or squared value of a quadrature $\hat x_\varphi$.
In the following, we show how such sign-free homodyne data can be used to characterize quantum states crucial for CVQIP.




\section{Quantum state certification}
\label{sec:dataprocessing}

This section introduces two methods for quantum certification of CV quantum states using sign-free quadrature measurement data. The first method used computationally efficient semidefinite programs tailored to sign-free homodyne data and enables the loss-tolerant tomographic reconstruction of a large class of CV quantum states (see section~\ref{sec:SDP}).
The second method combines homodyne estimators from~\cite{d2003quantum} adapted to sign-free data, with rigorous confidence intervals and allows for the direct estimation of the properties of CV quantum states (see section~\ref{sec:direct}). We illustrate applications of this versatile toolkit for loss-tolerant tomographic reconstruction and obtaining confidence intervals on important quantum features with experimental and numerical examples.


\subsection{Semidefinite Programming Reconstruction} \label{sec:SDP}

In this section, we discuss our full quantum state tomography protocol for parity-symmetric states, i.e., those states satisfying $\hat\Pi\rho\hat\Pi^\dag=\rho$, where $\hat\Pi=(-1)^{\hat N}$ is the photon-number parity operator. Equivalently, these states can be characterized by the property $\rho=\sum_{k=l\text{ mod }2}\rho_{kl}\ket k\!\bra l$, i.e.\ $\braket{k|\rho|l}=0$ if $k$ and $l$ have different parities. 

This category of states encompasses various notable examples such as photon-number states, binomial states~\cite{Binomial_code}, cat states~\cite{cat_code}, and GKP states~\cite{Gottesman2001}, which are leading candidates for bosonic quantum error correction.  
As discussed in Sec.~\ref {sec:expscheme}, a high-gain phase-sensitive OPA directly measures the sign-free quadratures, allowing us to reconstruct the probability distributions for $|x_\varphi|$ after performing a sufficiently large number of measurements for each phase-space quadrature. Applying the Born rule, one can subsequently determine the probability distribution for each phase-space quadrature. Mathematically, we get 
\begin{equation}
    \begin{aligned}
        P(|x_\varphi|) &= \text{Tr}[ \hat{\Pi}_\varphi\rho],\\
    \end{aligned}
    \label{eq: quad_distribution1}
\end{equation}
where $\hat{\Pi}_\varphi$ is the OPA measurement POVM, given in Eq.~\ref{eq:povm}. Writing the unknown density operator in the photon-number basis, Eq.~\ref{eq: quad_distribution1} simplifies to (see Appendix \ref{app:quadrature_distribution} for details):

\begin{equation}
    \begin{aligned}
        \label{eq:sdp_eqn}
        P(|x_\varphi|) &= \sum_{n,n' = 0}^{N_{\text{max}}} c_{n,n'} e^{-i\varphi(n'-n)}\Pi^{n',n}_{x_\varphi},\\
    \end{aligned}
\end{equation} 
where $N_{\text{max}}$ represents the Fock-space truncation due to finite energy limit. $\Pi^{n',n}_{x_\varphi} =H_{n}(x_\varphi)H_{n'}^*(x_\varphi)+H_n(-x_\varphi)H_{n'}^*(-x_\varphi)$ and $c_{n,n'} = \langle n|\rho| n'\rangle$, where $H_n$ are Hermite polynomials.  
By varying the pump phase, $\varphi$, one can obtain the $|x_\varphi|$ distributions for several phases and construct a system of linear equations that can be expressed in the matrix form ${\bf{P = \Pi M}}$ (see Appendix \ref{app:SDP}).

The only unknowns are the coefficients $c_{n,n'}$ describing the density matrix $\rho$, which one can estimate by inverting the equation. To circumvent the impact of inevitable experimental imperfections on the numerical stability of the inverted solution and the physicality for the density matrix $\rho$, we propose the following semidefinite programming (SDP) to run a convex quadratic optimization algorithm that minimizes the $\ell^2$-norm of the following equation, subject to physicality constraints:

\begin{align}
    \label{eq:SDP_program}
    {\text{Minimize}}\quad\quad&||{\bf{P-\Pi M}}||_2 + \gamma \Tr[\rho \hat{N}]\nonumber \\
    \text{Subject to}\quad&\rho\geq0\quad \text{and}\quad \text{Tr}[\rho] = 1,
\end{align}
where the above equation represents the compact form of Eq.~\ref{eq:sdp_eqn} as shown in Appendix \ref{app:SDP}.
{\bf M} corresponds to the Liouville vector representation of the unknown quantum state $\rho$, {\bf P} represents the Liouville vector representation of probability distributions ($P(x_{\varphi})$), ${\bf \Pi}$ represents the Liouville vector representation of the OPA POVM and $||.||_2$ denotes the $l_2$ norm defined as $||V||_2 = \sqrt{\sum_{i}|v_i|^2}$. The parameter $\gamma$ is an ad hoc parameter that penalizes larger photon numbers and is chosen to maximize the fidelity of the state reconstruction. The proposed inversion procedure is computationally efficient and yields a unique solution~\cite{boyd2004convex} while ensuring the physicality of the reconstructed state. 

While the proposed scheme allows full quantum state reconstruction of a parity-symmetric state, it can also be used to certify the non-classicality of an arbitrary state. A parity-symmetrized version of an arbitrary quantum state $\rho$ is:
\begin{equation}
    \tilde{\rho} = \frac{1}{2} \left( \hat{\Pi} \rho \hat{\Pi}^{\dagger} + \rho \right). 
\end{equation}
As a result, we have $\hat{\Pi} \tilde{\rho} \hat{\Pi}^{\dagger} = \tilde{\rho}$. Due to the parity symmetry of the POVM elements in Eq.~\eqref{eq:povm}, $\tilde{\rho}$ yields exactly the same outcome distribution as $\rho$ for the measurement described by Eq.~\eqref{eq:povm}. Since $\tilde{\rho}$ can be obtained by classical operations, i.e., a $\pi$ phase rotation applied with $1/2$ probability in $\rho$, any non-classicality detected in $\tilde{\rho}$ necessitates the non-classicality of $\rho$.

\begin{figure*}
    \centering
    \includegraphics[width = \textwidth]{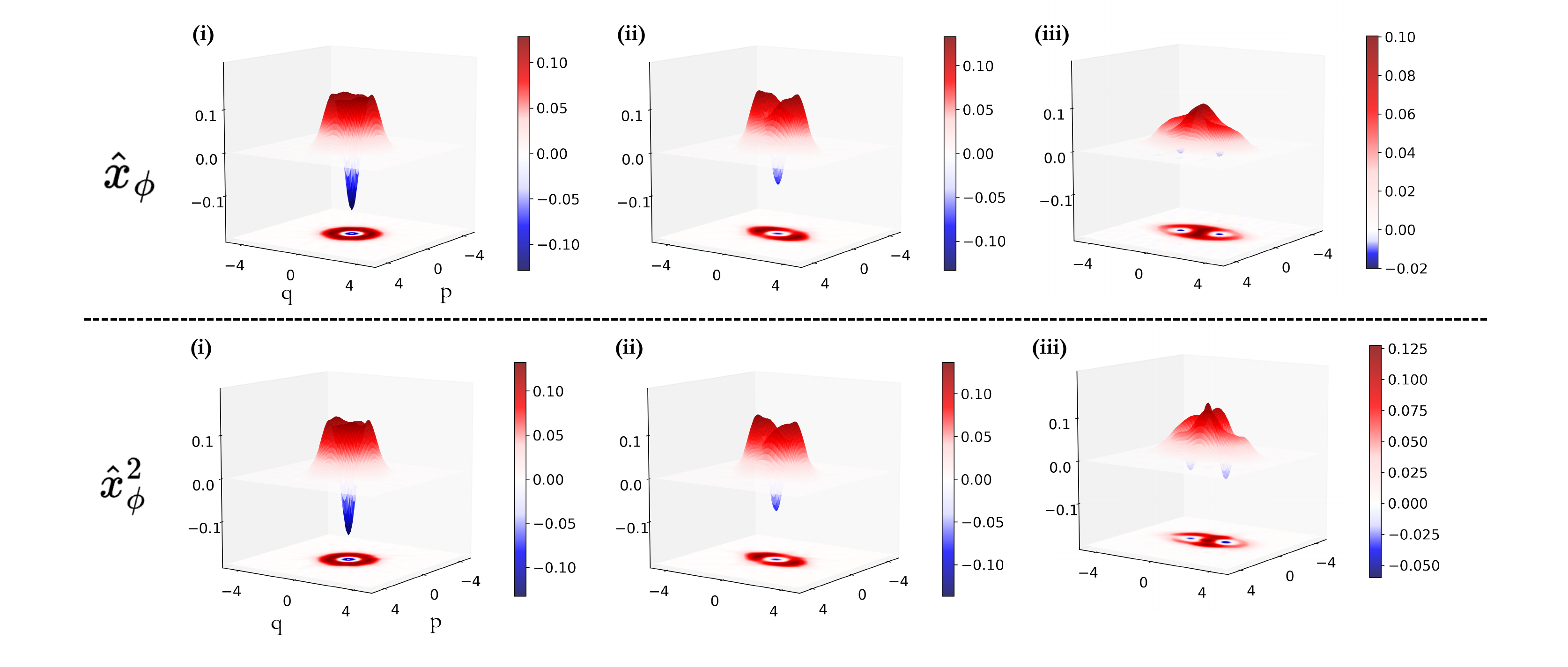}
    \caption{Comparison of quantum state reconstruction using conventional homodyne detection and the proposed OPA-based measurement protocol. Experimental homodyne data for the generalized quadrature $x_{\varphi}$ are taken from the experiments in Refs.~\cite{hanamura2025scalable_prx,kawasaki2024broadband,konno2024logical}. \textbf{Top row:} Wigner functions reconstructed for representative non-Gaussian states—single-photon, Schr\"odinger cat, and Gottesman--Kitaev--Preskill (GKP) states (left to right)—using balanced homodyne detection (BHD). The reconstruction employs the quadrature POVM $|x_{\phi}\rangle\langle x_{\phi}|$ together with the semidefinite program in Eq.~\ref{eq:SDP_program}. \textbf{Bottom row:} Wigner functions reconstructed from the proposed OPA-based measurement using the POVM defined in Eq.~\ref{eq:povm} and the corresponding absolute quadrature distribution $P(|x_{\varphi}|)$ derived from the same homodyne data. The close agreement between the two reconstructions demonstrates that faithful tomography of non-Gaussian states can be achieved using quadrature-power measurements without resolving the sign of the measured quadrature.}
    \label{fig:exp_data_sdp}
\end{figure*}

\begin{table*}
    \centering
    \begin{tabular}{ccccc}
       \toprule
    
        & Fid [SDP($x$)] & Fid [SDP($x^2$)] & Runtime ($x$, seconds) & Runtime ($x^2$, seconds)\\
           $\ket{1}$ (exp.) & 0.9984&0.9988& 1.29 &0.63\\
        cat (exp.) & 0.9961&0.9955& 0.53 & 0.36\\
        GKP (exp.) & 0.9963&0.9511& 1.88 &1.18\\
        $\ket{1}$ (sim.) & 0.9895&0.9896& 1.1 & 0.40\\
        cat (sim.) & 0.9870&0.9910 & 18.2 & 5.77\\
        GKP (sim.) & 0.9949&0.9958& 52.1 &15.3\\
     
        \hline
    \end{tabular}
    \caption{Reconstruction fidelity. For simulations, fidelity with the ideal state. For experiments, fidelity with MAXLIK($x$).}
    \label{tbl:fidelity}
\end{table*}
\subsection{Reconstruction for Experimental and Numerically Simulated Data}
\label{sec: verif}
To demonstrate the efficacy of our method for parity-symmetric states, we use experimental quadrature data obtained via conventional BHD for three types of quantum states: a single-photon state~\cite{hanamura2025scalable_prx}, a photon-subtracted squeezed (small-amplitude cat) state~\cite{kawasaki2024broadbandbroadband}, and a homodyne-bred approximate GKP state~\cite{konno2024logical}. The quadratures are experimentally sampled at six equally spaced phases between $0$ and $\pi$. We compare the reconstruction quality of conventional BHD with our proposed OPA-based tomography scheme, emulated using the absolute quadrature distributions $P(|x_{\varphi}|)$, which correspond to sign-insensitive homodyne measurements. The SDP programs are solved numerically using the open-source Python package CVXPY~\cite{cvxpy}. Similar convex optimization techniques have been employed in quantum detector tomography~\cite{Lundeen2009NP,Grandi_2017}.

\begin{figure*}
    \centering
    \includegraphics[width = \textwidth]{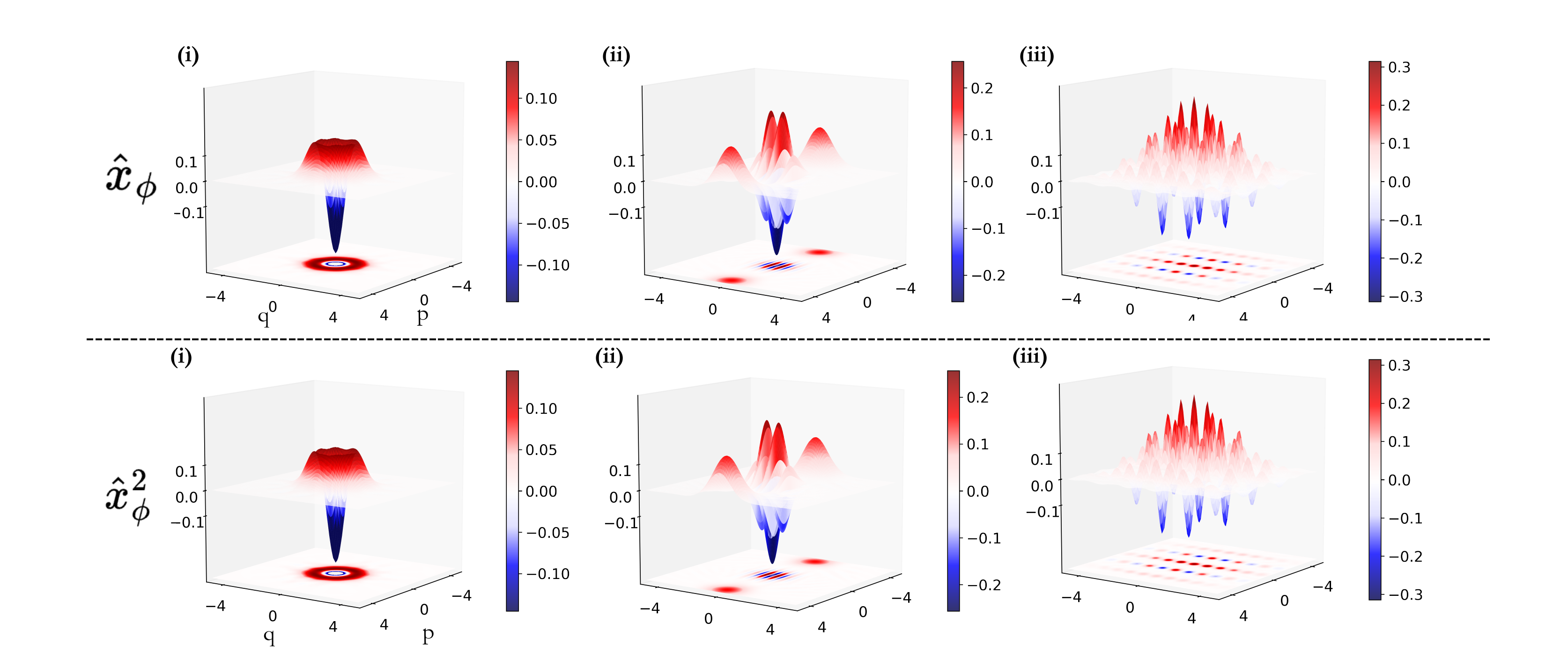}
    \caption{Comparison of quantum state reconstruction using homodyne and OPA-based measurements for representative non-Gaussian states. From left to right: single-photon, Schr\"odinger cat, and Gottesman--Kitaev--Preskill (GKP) states. \textbf{Top row:} Wigner functions reconstructed from conventional homodyne tomography using the generalized quadrature distribution $P(x_{\varphi})$ within the semidefinite-program (SDP) reconstruction framework. \textbf{Bottom row:} Wigner functions reconstructed from the proposed OPA-assisted protocol using the phase-insensitive quadrature-power distribution $P(|x_{\varphi}|)$ together with the corresponding OPA POVM in the same SDP framework. } 
    \label{fig:sim_states}
\end{figure*}

Fig.~\ref{fig:exp_data_sdp} displays the Wigner functions reconstructed using SDP for conventional BHD and our OPA-based tomography using experimental data for single-photon, cat, and GKP states.
The upper row shows reconstructions using the full $P(x_{\varphi})$ quadrature distributions, making the data tomographically complete with the POVM given as $\ket{x_{\varphi}}\bra{x_{\varphi}}$. This leads to $\Pi^{n',n}(x_{\varphi})=H_{n}(x_\varphi)H_{n'}^*(x_\varphi)$ as POVM elements in Eq.~\ref{eq:SDP_program}. Details are discussed in the Appendices~\ref{app:QST} and \ref{app:quadrature_distribution}. Similarly, the bottom row shows reconstructions using emulated OPA-based tomography, i.e., from $\hat{x}_\phi^2$ measurements. Here, we use the absolute quadrature distributions $P(|x_{\varphi}|)$, which correspond to a sign-free homodyne measurement. This leads to the POVM elements in Eq.~\ref{eq:povm} being $\Pi^{n',n}(x_{\varphi}) = H_n(x_\varphi) H_{n'}(x_\varphi) + H_n(-x_\varphi) H_{n'}(-x_\varphi)$. To benchmark the performance of the SDP-based method, we determined the fidelity $F(\rho, \sigma) = \left( \mathrm{Tr} \sqrt{ \sqrt{\rho} \, \sigma \, \sqrt{\rho} } \right)^2
$ of the reconstructed state with conventional BHD with the MaxLik algorithm state estimator. Table~\ref{tbl:fidelity} shows the fidelities and computation runtime on which the SDP solver converges (Apple M4 chip 12-core CPU, 16-core GPU, 24GB unified memory). We see that the SDP reconstruction shows near-unity for all states, both the complete quadrature data SDP($\hat{x}_\phi)$ and the sign-free data SDP($\hat{x}^2_\phi)$ with reduced run time for sign-free construction. The runtime advantage is more pronounced for complex states with larger photon numbers, as seen for the simulated GKP states, where the photon numbers were significant up to $N=35$ photons to truncate the Hilbert space. This is sufficient for a GKP state with stellar rank of 4.
The slight reduction in fidelity for experimental GKP with SDP($\hat{x}_\phi)$ may be attributed to the limited quadrature phases and noise in experiments~\cite{konno2024logical}.

Next, we perform simulations for single-photon ($\ket{1}$), cat ($\ket{\mathrm{cat}_{3}^{-}}$) and GKP ($\ket{\mathrm{GKP}_{0.3}}$) states. Quadrature data are sampled for 10, 10, and 40 phases, respectively. More details on simulation parameters are provided in Appendix \ref{app:Experimental parameters}, Table \ref{tbl:parameters}. The Wigner functions reconstructed using our SDP formalism for complete homodyne measurements are shown in Fig.~\ref{fig:sim_states} (upper panel), while the OPA-based measurements (or sign-free homodyne) using the absolute value of the quadratures are shown in Fig.~\ref {fig:sim_states} (lower panel). Similarly, we calculate the fidelity of the reconstructed state to the ideal state using simulated quadrature data. The results are shown in Table~\ref{tbl:fidelity}, achieving near-unity fidelity for all states using both complete quadrature data SDP($\hat{x}_\phi)$ and sign-free data SDP($\hat{x}^2_\phi)$ with reduced run time for sign-free construction. For the simulated GKP state, we observed a significant reduction in runtime using sign-free data, demonstrating the advantage of our method. 

In the following sections, we further introduce cost-effective methods for characterizing certain properties of target quantum states and for efficient estimation of linear functions.

\subsection{Direct Estimation of Linear Functions}
\label{sec:direct}

Identifying and quantifying the specific properties that enable quantum information processing, such as entanglement and nonclassicality, is essential for assessing the operational power of quantum states. In many settings, full density-matrix reconstruction is neither necessary nor easily measurable experimentally, particularly for quantum states in the mesoscopic photon-number regime with a large Hilbert space~\cite{yanagimoto2024mesoscopic, deng2024quantum}. 
Instead, resource-efficient certification protocols that directly probe the relevant quantum property offer a more practical alternative.

In CVQIP, various techniques to identify relevant quantum properties have been developed. For instance, Refs.~\cite{chabaud2020stellar,chabaud2021certification,fiuravsek2022efficient} provide witnesses for the stellar rank based on expectation values with simple observables, such as projectors onto a pure non-Gaussian target state. In this case, by determining the stellar robustness profile of the target state, that is, the set of maximum achievable fidelities with that target state for each fixed stellar rank, one can witness the stellar rank of any unknown state by estimating its fidelity with the target state using a few measurements. Similarly, Refs.~\cite{chabaud2021certification,chabaud2021witnessing} provide witnesses for Wigner negativity based on expectation values with simple observables, such as $\hat\Omega(\beta):=\hat D(\beta)(\ket1\!\bra1+\ket3\!\bra3+\ket5\!\bra5)\hat D(-\beta)$, where $\hat D(\beta)=e^{\beta\hat a^\dag-\beta^*\hat a}$ is a displacement operator, and $\beta$ is displacement amplitude. In that case, this observable is shown to have a threshold expectation value of $1/2$; that is, the maximum expected value of a Wigner-positive state is $1/2$. Using a few measurements, one can witness the Wigner negativity of any unknown state by estimating its expectation value with the operator $\hat\Omega(\beta)$. {Although such witnesses are, in principle, experimentally accessible, their practical implementation is highly susceptible to noise and loss for large photon number states. In all-photonic platforms in particular, the situation is further complicated by the limited range of photon-number-resolving detectors, as well as the difficulty of directly measuring projectors onto specific Fock subspaces, such as $(\ket{1}\bra{1} + \ket{3}\bra{3} + \ket{5}\bra{5})$.}


To overcome these challenges,  we introduce an efficient and reliable certification method that provides estimates with rigorous confidence intervals for a broad class of linear functionals of an experimentally prepared state, namely quantities of the form $\mathrm{Tr}(\hat{O}\rho)$. Our method applies to any parity-symmetric observable of the form

\begin{equation}
    \smash{\hat O}=\sum_{m=n\text{ mod }2}O_{mn}\ket m\!\bra n.
\end{equation}

Our scheme obtains samples from the proposed sign-free quadrature measurements using high-gain phase-sensitive amplifiers, rendering it robust to experimental loss and noise and removing reliance on photon-number-resolving detectors.
In particular, using these estimates and their associated confidence intervals requires no assumption about the state being characterized. The class of linear functions that can be reliably estimate with our method includes fidelities with eigenstates of the photon-number parity operator, namely states of the form
\begin{equation}
\ket{\phi}=\sum_{j\ge 0}\phi_j\ket{2j},
\qquad
\ket{\psi}=\sum_{j\ge 0}\psi_j\ket{2j+1}.
\end{equation}
This family encompasses a broad range of relevant non-Gaussian states, including Fock, cat, binomial, and GKP states, all critical to bosonic quantum error correction. Moreover, witnesses for the stellar rank~\cite{chabaud2021certification} or Wigner function negativity~\cite{chabaud2021witnessing} taking the form of fidelity estimates are available. As a result, our method provides a hardware-efficient, experimentally robust certification framework for the non-Gaussian properties of CV quantum states.

The estimators we introduce are linear combinations of homodyne tomography estimators~\cite{d2003quantum}.
For all $n,d \in \mathbb{N}$, $x \in \mathbb{R}$, $\varphi \in [0,\pi]$, and detection efficiency $\eta \le 1$, we define
\begin{align}
\mathcal{R}_{n,n+d}^{\eta}(x,\varphi)
&:=
e^{i d \left(\varphi+\frac{\pi}{2}\right)}
\sqrt{\frac{n!}{(n+d)!}}
\nonumber\\
&\quad\times
\int_{-\infty}^{+\infty}
dk\, |k|\,
e^{-\left(1-\frac{1}{2\eta}\right)k^2 - 2 i k x}
\, k^{d}
L_{n}^{(d)}(k^2),
\label{eq:homodyne_est}
\end{align}
where $L_n^{(d)}$ denotes the generalized Laguerre polynomial.

Let $\rho$ be a quantum state. In Appendix~\ref{app:direct}, we prove that for all $n,k \in \mathbb{N}$,
\begin{equation}
\label{eq:expectedvparity}
\rho^{\text{Estimated}}_{n+2k,n}
=
\mathbb{E}_{\varphi,\,|x_\varphi|}
\!\left[
\mathcal{R}_{n,n+2k}^{\eta}
\right],
\end{equation}
where the expected value is over the choice of a uniformly random angle $\varphi$ and the measurement outcomes $|x_\varphi|$ of a sign-free quadrature measurement of the state $\rho$ with overall quantum efficiency $\eta$. Using the hermiticity of the state, $\rho^\dagger = \rho$, we can therefore reconstruct all density-matrix elements $\rho_{m,n}$ with $m$ and $n$ of the same parity solely from sign-free quadrature measurement outcomes $(|x|,\varphi)$. 
Specifically, each element is obtained by taking the empirical mean of the estimator $\mathcal{R}_{n,m}^{\eta}$ over the collected data. We emphasize that \textit{sign-free quadrature measurements} are sufficient to fully characterize any parity-symmetric quantum state, allowing complete recovery of all experimentally accessible information.
The precision of the mean estimate is guaranteed by Hoeffding's inequality~\cite{hoeffding1963probability}. For an accuracy parameter $\epsilon>0$ and a failure probability $\delta>0$, and let $M$ denote an upper bound on the range of the estimator $\mathcal{R}_{n,n+2k}^{\eta}$. Then, for $N$ independent samples $(|x_1|,\varphi_1),\dots,(|x_N|,\varphi_N)$, Hoeffding's inequality implies that the estimated ${\rho}^{\text{estimated}}_{n+2k,n}$ satisfies
\begin{equation}
\Pr\!\left[
\bigl|
{\rho}^{\text{estimated}}_{n+2k,n}
-
\rho_{n+2k,n}
\bigr|
\ge \epsilon
\right]
\le \delta,
\end{equation}
provided that the number of measured samples 
\begin{equation}
N \ge 
\frac{2 M^2}{\epsilon^2}
\log\!\left(\frac{2}{\delta}\right).
\end{equation}

By linearity, the same strategy extends to arbitrary linear functionals of the state of the form
\(
\Tr(\hat O \rho)
\),
where
\(
\hat O=\sum_{m=n \, \mathrm{mod}\, 2} O_{mn} \ket m\!\bra n
\)
is a parity-symmetric operator. 
Such quantities can be estimated by computing the mean of the corresponding linear combination of estimators,
\begin{equation}
\sum_{m=n \, \mathrm{mod}\, 2}
O_{nm}\,
\mathcal{R}^{\eta}_{n,m}.
\label{eq:expec}
\end{equation}

With the union bound, using $N\ge\frac{2M^2}{\epsilon^2}\log\left(\frac{2K}\delta\right)$ samples, $K$ distinct linear features such as fidelities, density-matrix elements, or general expectation values can be estimated simultaneously with precision $\epsilon$ and confidence $1-\delta$. We stress that this scaling constitutes a worst-case analytical upper bound; in practice, substantially fewer samples typically suffice to achieve accurate estimation of linear functionals of the state.

\medskip

To validate our framework, we reconstruct the density matrix elementwise using sign-free quadrature measurement outcomes $(|x|,\varphi)$ and the estimators $\mathcal{R}^{\eta}_{n,m}$, as prescribed by Eq.~\ref{eq:expectedvparity}. 
This yields an elementwise estimate $\rho^{\mathrm{elementwise}}_{|x|}$ of the experimental state. We then compute the fidelity between the reconstructed state and the target state via SDP procedure described in the previous section. We use sign-free quadrature datasets $(|x|,\varphi)$ for both simulated states and experimental date of single-photon, cat, and GKP states. 
The resulting fidelity values are reported in Table~\ref{tbl:element_reconstruction}.

\begin{table}[h]
\centering
\small
\setlength{\tabcolsep}{4pt}
\begin{tabular}{lcccc}
\toprule
State 
& $\rho_{|x|}^{\mathrm{(simulated)}}$ 
& $\rho_{x}^{\mathrm{(simulated)}}$ 
& $\rho_{|x|}^{\mathrm{(experimental)}}$ 
& $\rho_{x}^{\mathrm{(experimental)}}$ \\

$\ket{1}$ 
& 0.987 & 0.991 
& 0.974 & 0.979 \\
Cat  
& 0.982 & 0.987 
& 0.980 & 0.976 \\
GKP 
& 0.989 & 0.992 
& 0.984 & 0.986 \\
\toprule
\end{tabular}
\caption{Fidelity between the elementwise reconstructed state 
$\rho^{\mathrm{elementwise}}_{|x|}$ and the target state, 
using either sign-free quadrature data ($|x|$) or full quadrature data ($x$), 
for simulated and experimental data.}
\label{tbl:element_reconstruction}
\end{table}

As an additional check, we compute the trace distance between the elementwise reconstruction $\rho^{\mathrm{elementwise}}_{|x|}$ and the SDP reconstruction $\rho^{\mathrm{SDP}}_{|x|}$. The trace distance $D = 1/2\left\|
\rho^{\mathrm{elementwise}}_{|x|}
-
\rho^{\mathrm{SDP}}_{|x|}
\right\|_1$ , where $\|\cdot\|_{1}$ denotes the trace norm,
$\|A\|_{1}
=
\Tr\!\left[\sqrt{A^{\dagger}A}\right]$. The trace distance is found to be $0.029$, $0.022$, and $0.015$ for the single-photon, cat, and GKP states, respectively. 
These small values confirm good agreement between the two reconstruction methods, with the corresponding near unity fidelities.



Finally, to avoid dealing with unbounded estimators, a target state $\ket\psi=\sum_{n=0}^{+\infty}\psi_n\ket n$ with unbounded support over the Fock basis for CV systems may be replaced by its truncated sub-normalized cutoff (experimentally determined by the largest photon number components present in the state) $\ket{\psi_{N_0}}=\sum_{n=0}^{N_0}\psi_n\ket n$. The fidelities with respect to $\ket{\psi}$ and its truncation $\ket{\psi_{N_0}}$ satisfy
\begin{equation}
    F(\rho,\ket{\psi_{N_0}})
    \;\le\;
    F(\rho,\ket{\psi})
    \;\le\;
    F(\rho,\ket{\psi_{N_0}})
    +
    D(\ket{\psi_{N_0}},\ket{\psi}), 
\end{equation}
where $F$ denotes the fidelity and $D$ the trace distance ($D=\sqrt{1-F}$ for pure states). Note that $D(\ket{\psi_{N_0}},\ket\psi)\rightarrow0$ when $N_0\rightarrow+\infty$.
The lower bound is obtained directly from $\rho$ being positive semidefinite, while the upper bound follows from the operational definition of the trace distance $D(\sigma,\tau)=\sup_{0\le P\le\mathds1}\text{Tr}[P(\sigma-\tau)]$~\cite{nielsen2002quantum}.

\begin{figure*}[!hbt]
	\begin{center}
		\includegraphics[width=\textwidth]{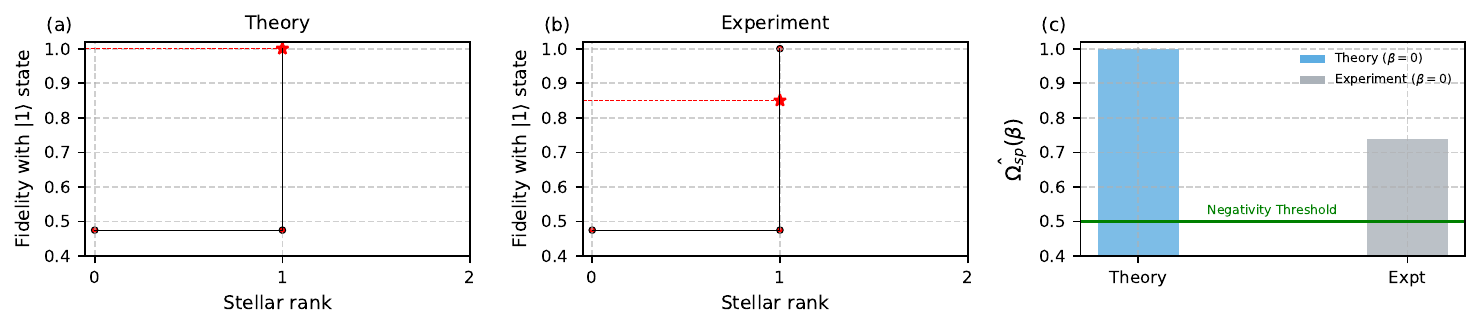}
		\caption{Certification of a single-photon state $\ket{1}$~\cite{hanamuru_prx} using using experimental homodyne tomography data and simulated sign-free quadrature data. (a) Stellar-rank witness for the simulated state obtained via direct fidelity estimation, together with the stellar robustness profile of a target single-photon state. (b) Stellar-rank witness for the experimental $\ket{1}$ state ~\cite{hanamuru_prx} obtained using the same procedure. In both panels, the red star denotes the measured fidelity mapped onto the stellar robustness profile, certifying stellar rank $1$. (c) Wigner-negativity witness for the simulated and experimental states using the expectation value of the operator 
$\Omega_{\mathrm{sp}}:=\hat{D}(\beta)\left(\ket{1}\!\bra{1}+\ket{3}\!\bra{3}+\ket{5}\!\bra{5}\right)\hat{D}(-\beta)$. For $\beta=0$, the measured expectation values exceed the threshold of $0.5$, certifying Wigner negativity.}
		\label{fig:examplesp}
	\end{center}
 
\end{figure*}

As a result, we can reliably use the fidelity with the cutoff state as a witness for the true fidelity. This allows us to retrieve information on an experimental state 
without assuming \textit{a priori} that the state has bounded Fock support.
Although finite-energy support is typically a reasonable assumption in experimental scenarios and is often implicitly adopted, it is preferable to avoid it in the context of certification of non-Gaussian properties. Indeed, an artificial cutoff may itself induce non-Gaussian features in an otherwise Gaussian state (e.g., a coherent state or a squeezed state distinct from the vacuum), potentially leading to misleading conclusions.

Let us illustrate this procedure with the example of an even cat state. The state is

\begin{equation}
    \begin{aligned}
        \ket{\text{cat}_\alpha^+}&:=\frac1{\sqrt{2(1+e^{-2|\alpha|^2})}}(\ket\alpha+\ket{-\alpha})\\
        &=\frac1{\sqrt{\cosh(|\alpha|^2)}}\sum_{n=0}^{+\infty}\frac{\alpha^{2n}}{\sqrt{(2n)!}}\ket{2n},
    \end{aligned}
\end{equation}
where $\alpha\in\mathbb C$ is the amplitude of the coherent state.
Let $\ket{\mathrm{cat}^{+}_{\alpha,N_0}}$ denote the renormalized even cat state truncated at the Fock number $N_0$. 
Then, for an arbitrary density operator 
\(
\rho=\sum_{k,l=0}^{\infty}\rho_{k,l}\ket{k}\!\bra{l},
\)
the fidelity with respect to $\ket{\mathrm{cat}^{+}_{\alpha,N_0}}$ can be expressed as

\begin{equation}
    \begin{aligned}
        F(\rho,\ket{\text{cat}_\alpha^+})&\le F(\rho,\ket{\text{cat}^+_{\alpha,N_0}})+D(\ket{\text{cat}^+_{\alpha,N_0}},\ket{\text{cat}^+_{\alpha}})\\
        &\le F(\rho,\ket{\text{cat}^+_{\alpha,N_0}})+\sqrt{\frac{|\alpha|^{4N_0+2}e^{|\alpha|^2}}{(2N_0+1)!}},
    \end{aligned}
\end{equation}
where the second line is obtained by bounding the remainder of the Taylor series of the $\cosh$ function as derived in Appendix \ref{app:Fidelity_bd}.  
This ensures the quality of the fidelity witness $F(\rho,\ket{\text{cat}^+_{\alpha,N_0}})$ for $N_0$ large enough. For instance, for $|\alpha|=1$ and $N=6$ the error term is at most $\mathcal{O}(10^{-5})$. Then, up to normalization, the corresponding estimator for the fidelity with that cutoff state is
\begin{equation}
f_{\text{cat},N_0}^\eta(x,\varphi;\alpha)\propto\sum_{m,n=0}^{N_0}\frac{\alpha^{2n}\alpha^{*2m}}{\sqrt{(2m)!(2n)!}}\mathcal R^\eta_{2n,2m}(x,\varphi).
\end{equation}

\medskip

Next, we apply our direct fidelity estimation framework to certify both the stellar rank and the Wigner negativity of various quantum states, using simulated and experimental homodyne datasets, which we turn into sign-free homodyne by taking the absolute value. 
In particular, we consider three representative non-Gaussian states: a single-photon state $\ket{1}$ ~\cite{hanamuru_prx}, a photon-subtracted squeezed state ~\cite{Kawasaki2024}, and a qubit GKP state~\cite{konno2024logical} with envelope parameter $\Delta=0.3$ (we refer to Appendix~\ref{eq:approxGKP} for the formal definition).

Firstly, to certify the stellar rank of both experimental and simulated states, we reconstruct the density matrix $\rho_{\mathrm{SDP}}$ using the proposed SDP approach described in the previous section. We then construct the \emph{stellar rank profile} \cite{chabaud2020certification} of a chosen non-Gaussian target state (see Appendix~\ref{app:profiles}), defined as the set of maximal achievable fidelities with that target state for states of fixed stellar rank. Next, we compute the fidelity between the reconstructed state and the target state and compare it with this stellar rank profile, which serves as a lookup table for the lower bound on the stellar rank of the unknown state.

\begin{figure*}[!hbt]
	\begin{center}
		\includegraphics[width=\textwidth]{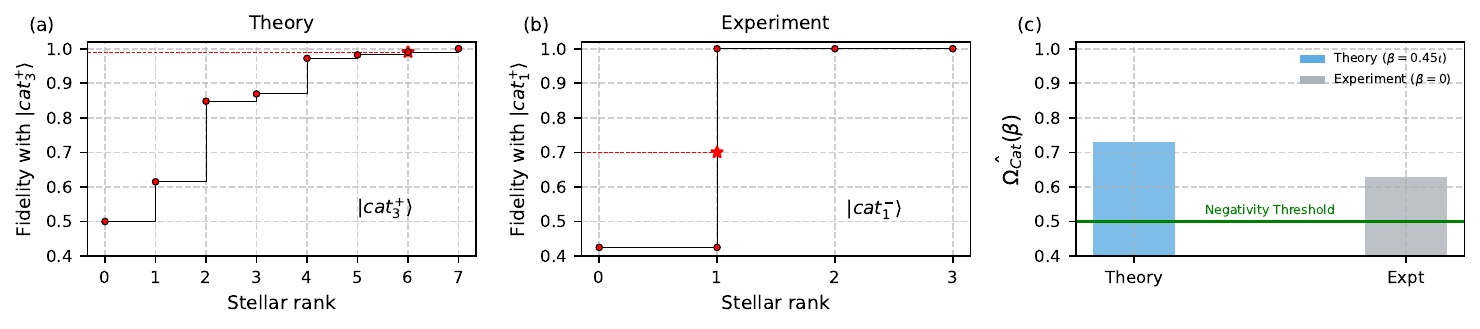}
		\caption{Certification of cat states using the homodyne tomography data from experiments in Ref.~\cite{Kawasaki2024} and simulated sign-free quadrature measurement data. (a) Stellar rank witnessing of simulated sign-free $\ket{\text{cat}^+_{3}}$ state using our direct fidelity estimation method together with the stellar robustness profile of a target $\ket{\text{cat}^+_{3}}$ state. (b) Stellar rank witnessing of experimental small amplitude cat state~\cite{Kawasaki2024} using our direct fidelity estimation method together with the stellar robustness profile of a target $\ket{\text{cat}^-_1}$ state. (c) Wigner negativity witnessing of the simulated and experimental small amplitude cat state using our direct fidelity estimation method together with the Wigner negativity witness $\Omega_\text{cat}:=\hat D(\beta)(\ket1\!\bra1+\ket3\!\bra3+\ket5\!\bra5)\hat D(-\beta)$, with $\beta=0.45i$ and $\beta=0$. }
		\label{fig:examplecat}
	\end{center}
 
\end{figure*}

\begin{figure*}[!hbt]
	\begin{center}
		\includegraphics[width=\textwidth]{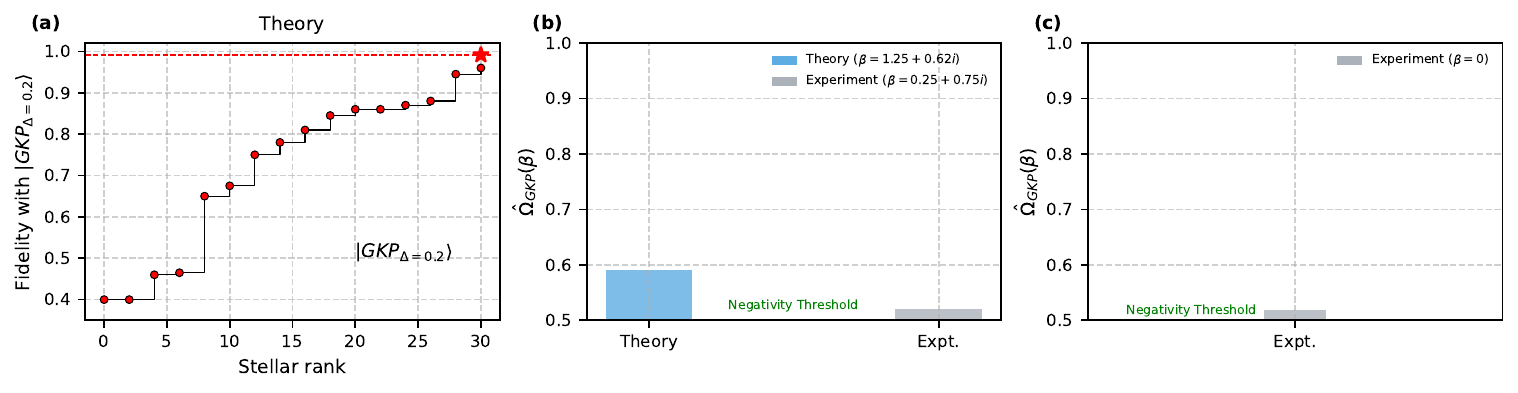}
		\caption{Certification of a $\ket{\mathrm{GKP}}$ state using experimental homodyne tomography data~\cite{konno2024logical}, \cite{Larsen2025} and simulated sign-free quadrature data. (a) Stellar rank witnessing of simulated sign-free $\ket{\mathrm{GKP}_{\Delta=0.2}}$ state using our direct fidelity estimation method together with the stellar robustness profile of a target $\ket{\mathrm{GKP}_{\Delta=0.2}}$ state. (b) Wigner negativity witnessing of the simulated and experimental state (Ref.~\cite{konno2024logical}) using our direct fidelity estimation method together with the Wigner negativity witness $\Omega_\text{GKP}:=\hat D(\beta)(\ket1\!\bra1+\ket3\!\bra3+\ket5\!\bra5)\hat D(-\beta)$, with $\beta=1.25+0.62i$ and $\beta=0.25+0.75i$ respectively. (c) Wigner negativity witnessing of the experimental Xanadu's state \cite{Larsen2025} with the Wigner negativity witness $\Omega_\text{GKP}$ at $\beta=0$.}
		\label{fig:exampleGKP}
	\end{center}
\end{figure*}

Furthermore, we witness the nonclassicality of the considered states by determining the negativity of their Wigner functions, a hallmark of quantumness that cannot arise for classical states~\cite{mari2012positive}. We detect Wigner negativity using the witnesses introduced in Refs.~\cite{chabaud2021certification,chabaud2021witnessing}. Specifically, we estimate the expectation value of the operator
$
\hat{\Omega}(\beta):=\hat{D}(\beta)\left(\ket{1}\!\bra{1}+\ket{3}\!\bra{3}+\ket{5}\!\bra{5}\right)\hat{D}(-\beta)$. We emphasize that our approach does not require full reconstruction of the density matrix $\rho$. Instead, linear fidelities can be directly estimated from quadrature data. This significantly reduces the experimental and computational overhead, particularly for quantum states with mesoscopic photon numbers, by eliminating the need for full quantum state tomography and allowing the measured quadrature statistics to be used directly. If, for any displacement $\beta$, the expectation value satisfies $\langle \hat{\Omega}(\beta) \rangle > 0.5$, the state is certified to exhibit Wigner negativity at the corresponding point $\beta$ in phase space.

As a special case for $\beta=0$, the witness operator becomes parity symmetric, $
\hat{\Omega}(0):=\ket{1}\!\bra{1}+\ket{3}\!\bra{3}+\ket{5}\!\bra{5}$. 
In this case, the expectation value of the unknown state can be directly estimated from \emph{sign-free} homodyne quadrature data, using the homodyne estimators introduced in Eq.~\ref{eq:homodyne_est} along with Eq.~\ref{eq:expec}. If the measured value satisfies $\langle \hat{\Omega}(0)\rangle > 0.5$, the state is certified to exhibit Wigner negativity at the origin in phase space.

For any nonzero displacement $\beta$, the expectation value of the witness operator $\hat{\Omega}(\beta)$
can be interpreted as estimating the parity-symmetric witness $\hat{\Omega}(0)$ on the displaced state $
\rho_\beta=\hat{D}(-\beta)\rho\hat{D}(\beta)$. 
Therefore, measuring $\langle \hat{\Omega}(\beta)\rangle$ for the state $\rho$ is equivalent to probing $\langle \hat{\Omega}(0)\rangle$ for the displaced state $\rho_\beta$. If Wigner negativity is detected at the origin for $\rho_\beta$, this directly implies that the original state $\rho$ possesses Wigner negativity at the phase-space point $\beta$. This follows from the fact that displacement is a Gaussian unitary operation that simply translates the Wigner function in phase space without distorting its shape or introducing negativity~\cite{nehra2019state}.

We next apply the above methods to characterize the stellar rank and Wigner negativity of both simulated and experimental single-photon states shown in Fig.~\ref{fig:examplesp}, using only sign-free homodyne quadrature data. Figures~\ref{fig:examplesp}(a,b) present the stellar-rank witnesses obtained by estimating the fidelity of the reconstructed state with a target single-photon state and comparing it with the corresponding stellar robustness profile (see Ref.~\cite{codestellarprofile} for the numerical implementation). The red star marks the measured fidelity placed on the stellar profile, from which we certify a stellar rank of $1$ for both the simulated and experimental states. Figure~\ref{fig:examplesp}c shows the Wigner-negativity witness obtained using the operator
$\hat{\Omega}_{\mathrm{sp}}:=\hat{D}(\beta)\left(\ket{1}\!\bra{1}+\ket{3}\!\bra{3}+\ket{5}\!\bra{5}\right)\hat{D}(-\beta)$. 
For $\beta=0$, the measured expectation value exceeds the threshold of $0.5$ for both states, thereby certifying the nonclassical nature.

Likewise, Fig.~\ref{fig:examplecat} (a, b) show the stellar rank witnesses of a simulated $\ket{\mathrm{cat}_{3}^{+}}$ state and experimental data from  small amplitude cat state (photon subtracted squeezed vacuum state) by estimating its fidelity with a target non-Gaussian $\ket{\mathrm{cat}^+_{3}}$, $\ket{\mathrm{cat}^-_{1}}$ states respectively together with their stellar robustness profiles. We then map the fidelity on the stellar profiles and witness a stellar rank of 2 for both states. Fig.~\ref{fig:examplecat}c shows the Wigner negativity witness using $\Omega_\text{cat}:=\hat D(\beta)(\ket1\!\bra1+\ket3\!\bra3+\ket5\!\bra5)\hat D(-\beta)$. As expected, we witness Wigner negativity ($>0.5$) for both states at $\beta=0.45i$ for the simulated state, and $\beta=0$ for the photon-subtracted squeezed state. Note that the simulated data correspond to an even-photon-parity cat state, whose Wigner negativity appears at a displaced phase-space point $\beta = 0.45 i$. In contrast, the experimentally generated state has odd photon-number parity, resulting in Wigner negativity centered near $\beta = 0$.

Fig.~\ref{fig:exampleGKP}a show the stellar rank witnesses of a simulated $\ket{\mathrm{GKP}_{\Delta=0.2}}$ state by estimating its fidelity with a target non-Gaussian $\ket{\mathrm{GKP}_{\Delta=0.2}}$ state together with its stellar robustness profile. For the experimental GKP states~\cite{konno2024logical,Larsen2025} we could not witness a stellar rank, as the fidelity of the experimental state with the targeted pure GKP state is low, given that the experimental states are highly mixed (see Appendix~\ref{app:xanadu} for a discussion). On the other hand, Fig.~\ref{fig:exampleGKP}b shows the Wigner negativity witness using $\Omega_\text{GKP}:=\hat D(\beta)(\ket1\!\bra1+\ket3\!\bra3+\ket5\!\bra5)\hat D(-\beta)$. We witness the Wigner negativity ($>0.5$) for both the states at $\beta=1.25+0.62i$, $\beta=0.2+0.75i$ respectively (where the homodyne data is translated before taking the absolute value). Fig.~\ref{fig:exampleGKP}c shows the Wigner negativity witness for Xanadu's GKP state~\cite{Larsen2025} using $\Omega_\text{GKP}:=\hat D(\beta)(\ket1\!\bra1+\ket3\!\bra3+\ket5\!\bra5)\hat D(-\beta)$. We witness the Wigner negativity ($>0.5$) at $\beta=0$. Sign-free quadrature measurements can thus reliably estimate the stellar rank and Wigner negativity of parity-symmetric states with rigorous confidence intervals, although we note that the fidelity-based witnesses used for stellar rank are too stringent for witnessing the stellar of highly mixed experimental GKP-like states.


\section{Conclusion and perspectives}
\label{sec:conclusion}


We have introduced a unified framework for the characterization and certification of non-Gaussian states based on high-gain phase-sensitive optical parametric amplification and power measurements. By combining all-optical, loss-tolerant measurements with computationally efficient semidefinite-program reconstruction, our approach enables full quantum state tomography of parity-symmetric continuous-variable states from quadrature power statistics while substantially relaxing detector-efficiency requirements and extending the accessible measurement bandwidth in traditional balanced homodyne detection. \\
Building upon this measurement protocol, we developed rigorous certification methods for non-Gaussian quantum states directly from sign-free quadrature data. In particular, we demonstrated the reconstruction and certification of representative states for simulated and experimental data, including single-photon, cat, and GKP states, through full-quantum state tomography, stellar-rank witnesses, and Wigner-negativity observables. These results establish that sign-free quadrature measurements contain sufficient information not only for accurate state reconstruction for parity-symmetric states but also for the certification of nonclassical resources relevant to continuous-variable quantum information processing.\\
Some of the certification methods introduced in this work can be readily generalized to the multimode setting. Although the number of samples required for a complete tomography inevitably scales exponentially in the number of modes, efficient methods for direct estimation of multimode CV quantum features are available based on balanced homodyne and heterodyne data~\cite{chabaud2021efficient,wu2021efficient,upreti2024efficient}. As we have shown, these methods can be adapted to sign-free data, yielding direct estimation methods that are efficient for a large class of CV multimode quantum states.\\
Similarly, we expect that our methods can be adapted beyond quantum state certification, e.g.\ for quantum process tomography using sign-free quadrature data, based on high-gain OPAs. Regarding the certification of the stellar rank, our analysis shows that linear witnesses, which are based on estimating the expectation values of specific observables, are not well-suited for highly mixed states. This motivates the development of more efficient, non-linear witnesses for stellar rank, which use multiple copies of the state, for instance via witness expansion of existing linear witnesses~\cite{tang2026witness}.\\

Finally, while we validated the proposed framework using experimental homodyne data, a natural next step is to implement it beyond Gaussian states~\cite{williams2025ultrafast}.  More broadly, our work highlights an interesting research direction: exploiting measurements that, while not tomographically complete, are very practical to implement experimentally. We leave pursuing this research direction for future work. We anticipate that this framework will provide a scalable route toward the verification and benchmarking of increasingly complex bosonic quantum processors, quantum communication networks, and quantum sensing platforms in both the optical and microwave domains.\\


\section*{Acknowledgements}

We thank Akito Kawasaki, Akira Furusawa, Young-Sik Ra, Valentina Parigi and Nicolas Treps for valuable discussions.  
U.C.~acknowledges funding provided by the Institute for Quantum Information and Matter, an NSF Physics Frontiers Center (NSF Grant PHY-1733907) and from the European Union’s Horizon Europe Framework Programme (EIC Pathfinder Challenge project Veriqub) under Grant Agreement No.\ 101114899. M.G.~acknowledges funding from the Hybrid Quantum Initiative (HQI) supported by France 2030 under ANR
grant ANR-22-PNCQ-0002. RN gratefully acknowledges support from the ECE Department at UMass Amherst.


\bibliographystyle{apsrev4-2}
\bibliography{biblio,biblio2,Quantum}


\appendix

\newpage
\onecolumngrid

\section{Experimental scheme}
\label{app:QST}


To derive the POVM for the OPA scheme, we start from Eq.~\ref{eq:I_distribution}. 
Substituting $\hat{I} = G \hat{x}_\varphi^{2}$ and $I=Gx_\varphi^2$ into the characteristic function $\chi(\lambda)$ and 
expanding the density operator in the eigenbasis of 
$\hat{x}_\varphi$, we obtain
\begin{align}
P(I)=P(x_\varphi)
&= \int_{-\infty}^{\infty} \frac{d\lambda}{2\pi} 
   e^{-i\lambda I}\, \chi(\lambda) \notag \\
&= \frac{1}{G} \int_{-\infty}^{\infty} \frac{dz}{2\pi}
   e^{-iz x_\varphi^{2}}
   \operatorname{Tr}\!\left[ \rho\, e^{iz \hat{x}_\varphi^{2}} \right] .
\end{align}
Expressing the density operator as
\begin{equation}
\rho
=
\int dx_\varphi' dx_\varphi''\,
\rho(x_\varphi',x_\varphi'')
\, |x_\varphi'\rangle \langle x_\varphi''|. 
\end{equation}
The probability distribution becomes
\begin{equation}
P(x_\varphi)=
\frac{1}{G} \int_{-\infty}^{\infty} \frac{dz}{2\pi}
e^{-iz x_\varphi^{2}}
\int dx_\varphi' dx_\varphi''\,
\rho(x_\varphi',x_\varphi'')
\operatorname{Tr}\!\left[
|x_\varphi'\rangle \langle x_\varphi''|
\, e^{iz \hat{x}_\varphi^{2}}
\right].
\end{equation}
Using the trace identity $\operatorname{Tr}\!\left[ |x\rangle \langle x'| \hat{O} \right]
=
\langle x'| \hat{O} |x\rangle$,
we obtain
\begin{equation}
    \begin{aligned}
P(x_\varphi)
&=
\frac{1}{G}
\int_{-\infty}^{\infty} \frac{dz}{2\pi}
e^{-iz x_\varphi^{2}}
\int dx_\varphi' dx_\varphi''\,
\rho(x_\varphi',x_\varphi'')
\,
\langle x_\varphi'' |
e^{iz \hat{x}_\varphi^{2}}
| x_\varphi' \rangle\\
&=\int  dx_{\varphi}'dx_{\varphi}'' \rho(x_{\varphi}',x_{\varphi}'')\frac{1}{G}\int \frac{dz}{2\pi}\bra{x_{\varphi}''}e^{iz(\hat{x}_\varphi^2-{x}_{\varphi}^2)} \ket{x_{\varphi}'}\\
&=\int  dx_{\varphi}'dx_{\varphi}'' \rho(x_{\varphi}',x_{\varphi}'')\bra{x_{\varphi}''}\left(\frac{1}{G}\int \frac{dz}{2\pi}e^{iz(\hat{x}_\varphi^2-{x}_{\varphi}^2)}\right) \ket{x_{\varphi}'}.
    \end{aligned}
\end{equation}

We now compare this expression with the probability distribution of $x_{\varphi}$ given by the Born rule, where we again write $\rho$ in the eigenvector basis of $\hat{x}_{\varphi}$:
\begin{equation}
    \begin{aligned}
        P(x_{\varphi})&= \text{Tr}[\rho \Pi_{x_\varphi}]\\
        &=\text{Tr}\left[\int dx_{\varphi}'dx_{\varphi}'' \rho(x_{\varphi}',x_{\varphi}'')\ket{x_{\varphi}'}\bra{x_{\varphi}''}\Pi_{x_\varphi} \right]\\
        &=\int dx_{\varphi}'dx_{\varphi}'' \rho(x_{\varphi}',x_{\varphi}'') \bra{x_{\varphi}''}\Pi_{x_\varphi}\ket{x_{\varphi}'},
    \end{aligned}
\end{equation}
which gives
\begin{equation}
\Pi_{x_\varphi}=\int dq_\varphi\left(\frac{1}{G}\int \frac{dz}{2\pi}e^{iz(q_\varphi^2-{x}_{\varphi}^2)}\right)\ket{q_\varphi}\!\bra{q_\varphi}.
\end{equation}
Now we have $\int \frac{dz}{2\pi}e^{iz(a^2-b^2)}=\delta(a^2-b^2)=\frac{\delta(a+b)+\delta(a-b)}{2|b|}$, so

\begin{equation}
    \begin{aligned}
     \Pi_{x_\varphi}&=\frac{1}{G}\int dq_\varphi\delta(q_{\varphi}^2 - x_{\varphi}^2)\ket{q_\varphi}\!\bra{q_\varphi}\\
        &=\frac{1}{G}\int dq_\varphi\left(\frac{\delta(q_{\varphi}-x_{\varphi})+\delta(q_{\varphi}+x_{\varphi})}{2|x_{\varphi}|}\right)\ket{q_\varphi}\!\bra{q_\varphi}\\
        &=\frac1{2G|x_\varphi|}(\ket{x_{\varphi}}\bra{x_{\varphi}} + \ket{-x_{\varphi}}\bra{-x_{\varphi}}).
    \end{aligned}
\end{equation}

\begin{figure*}
    \centering
    \includegraphics[height=6cm, width = \textwidth]{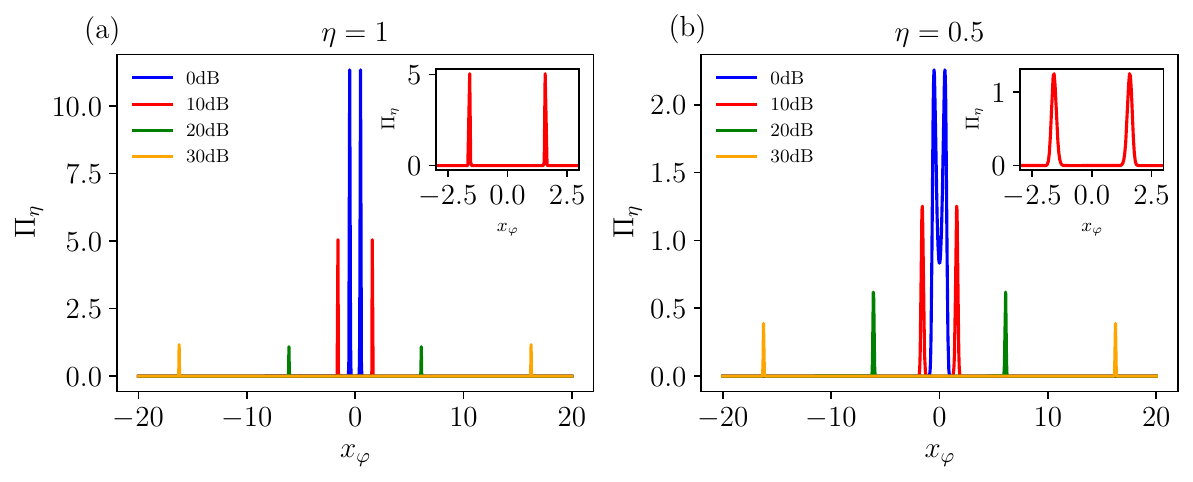}
    \caption{POVM plots for inefficient photodetection with efficiency $\eta$ and function of gain parameter $G=e^{r}$, where gain in $\mathrm{dB}$ is given by $10\log_{10}(e^{2r})$. (a) POVM for ideal photodetection ($\eta=1$) at different amplifier gains. The inset shows a magnified view of the POVM corresponding to a $10\,\mathrm{dB}$ gain. (b) POVM distributions for a detector with $\eta=0.5$ at various amplifier gains. The inset highlights the corresponding POVM for a $10\,\mathrm{dB}$ gain.}

    \label{fig:inefficient_opa}
\end{figure*}

Now, let us consider the case of inefficient photodetection where the detection efficiency of the photodetector is $\eta$. Now, it has been shown in \cite{d1999parametric} that an inefficient photodetector with efficiency $\eta$ can be replaced by a beam-splitter of transmittivity $\tau = \eta$, followed by a perfect detector ($\eta=1$) and the modified quadrature $X_{\varphi}^{\eta}$ is given by:
\begin{equation}
    \begin{aligned}
        X_{\varphi}^{\eta} \to X_{\varphi} + \sqrt{\frac{1-\eta}{2\eta}}v_{\varphi},
    \end{aligned}
\end{equation} 
where the the generalised vacuum quadrature $v_{\varphi}=v^{\dagger}e^{i\varphi}+ve^{-i\varphi}$ is the added noise from the open port of the beam splitter. We follow a similar approach here for our experimental setup. Here the perfect photodetector after the OPA can be replaced by a beam splitter of transmitivity $\eta$ followed by a perfect photodetector. The inefficient POVM is found by replacing $X_{\varphi}$ by $X_{\varphi}^{\eta}$ and tracing out the vacuum mode as follows:
\begin{equation}
    \label{eq:Inefficient_povm}
    \begin{aligned}
        \bra{x_\varphi'}\Pi^{\eta}_{x_\varphi}\ket{x_\varphi'}&=\frac{1}{2\pi}\bra{0}\int d\lambda e^{i\lambda(G\left(x_{\varphi}^{\eta}\right)^2-G(x'_{\varphi})^{2})}\ket{0}\\
        &=\frac{1}{2\pi}\int d\lambda e^{i\lambda(Gx_{\varphi}^2-Gx'^{2}_{\varphi})}\bra{0}e^{\sqrt{2}G\Delta v_{\varphi}+\frac{G}{2}\Delta^2 v^2}\ket{0}\\
        &=\frac{1}{2\pi}\int d\lambda e^{i\lambda(Gx_{\varphi}^2-Gx'^{2}_{\varphi})}\bra{0}e^{\sqrt{2}G\Delta v_{\varphi}}\ket{0}\\
        &=\frac{1}{2\pi}\int d\lambda e^{i\lambda(Gx_{\varphi}^2-Gx'^{2}_{\varphi})}e^{\frac{-\lambda^2 G\Delta^2}{4}}\\
        &= \frac{1}{\Delta\sqrt{G\pi}} \exp\left[\frac{-G(x_{\varphi}^2-x'^{2}_{\varphi})^2}{\Delta^2}\right],
    \end{aligned}
\end{equation}
where $\Delta^2=\frac{1-\eta}{4\eta}$, and $G=e^{2r}$.

Setting $\sigma^{2}=G^{-1}\Delta^2=e^{-2r}\Delta^2$, the above equation is a distribution with peaks at $y'=x_{\varphi}, -x_{\varphi}.$ The noise from the vacuum convolutes the POVM and is captured by the width parameter $\sigma^2$. This noise can be made arbitrarily small in the limit $r\to \infty$ and the POVM becomes:
\begin{equation}
        \Pi^{\eta}\propto\ket{x_{\varphi}}\bra{x_{\varphi}} + \ket{-x_{\varphi}}\bra{-x_{\varphi}},
\end{equation}
Up to a rescaling factor, this is the OPA POVM for an ideal measurement. Hence, our experimental setup is protected from the vacuum noise in the high gain limit.

Fig.\ \ref{fig:inefficient_opa} shows the plots for inefficient POVM for the OPA in Eq.~\ref{eq:Inefficient_povm} for different detector efficiency ($\eta$) and as a function of different gains. Fig.\ \ref{fig:inefficient_opa}(a) shows that for a perfect detector the POVM is equivalent to two sharply peaked $\delta$ functions at $x_{\varphi}, -x_{\varphi}$ as the width ($G^{-1}\Delta^2$) for $\eta \to 1$ gives $\Delta \to 0$. For larger gains we see a scaling of $x_{\varphi}$ as expected from Eq.~\ref{eq:Inefficient_povm}. Fig.~\ref{fig:inefficient_opa}(b) shows the POVM for detector with efficiency $\eta=0.5$ for different gains. The peaks at $x_{\varphi}, -x_{\varphi}$ have a non-zero width ($G^{-1}\Delta^2$) due to the convolution from vacuum noise. For no gain we see that both peaks have merged due to noise unlike the sharp peaks for $\eta=1$. The inset for $\eta=0.5$ shows the zoomed in POVM for $10\mathrm{dB}$ gain and we see that there is a non-zero width unlike for the POVM for $\eta=1$ with same gain.

\section{Experimental and simulation parameters}
\label{app:Experimental parameters}

Table~\ref{tbl:parameters} summarizes the parameters for each state tomography. $N_{max}$ is the cutoff dimension of the density matrix; $N_{bins}$ is the number of bins when making histograms; the range of the histogram is $(-L,L)$ for homodyne tomography and $(0,L)$ for OPA tomography; $N_{phases}$ is the number of phases in the range $(0,2\pi)$; $N_{points}$ is the number of measured points for each phase.

\begin{table}[h]

    \centering
    \begin{tabular}{cccccc}
       \toprule
       
        & $N_{max}$ & $N_{bins}$ & $L$ & $N_{phases}$ &  $N_{points}$ \\

  $\ket{1}$ (exp.) & 35 & 40&5 & 6 & 3333\\
        cat (exp.) & 20 & 40&5 & 6 & 5000\\
        GKP (exp.) & 35&40&5&6&19610\\
        $\ket{1}$ (sim.) & 20&40&5&10&5000\\
        cat (sim.) & 35&200&8&10&10000\\
        GKP (sim.) & 35&200&8&40&20000\\
       
 \hline
    \end{tabular}
    \caption{Parameters used in SDP reconstruction.}
    \label{tbl:parameters}
\end{table}

\section{Probability distribution of generalized quadrature}
\label{app:quadrature_distribution}
The probability distribution of the generalized quadrature $x_{\varphi}$ can be written as follows using Born's rule:
\begin{equation}
    \begin{aligned}
        P(|x_\varphi|) &= \text{Tr}[ \hat{\Pi}_\varphi\rho].\\
        &\approx\text{Tr}\!\left[\frac12(|x_\varphi\rangle \langle x_\varphi| + |-x_\varphi\rangle \langle- x_\varphi|) \rho\right].
    \end{aligned}
    \label{eq: quad_distribution_1}
\end{equation}
We write the generalized quadrature $x_{\varphi}$ and $\rho$ in the photon number basis as follows:
\begin{equation}
\label{eq:fock_basis1}
    \begin{aligned}
        |{x}_\varphi\rangle &= \frac{e^{-x_\varphi^2/2}}{\pi^{1/4}}\sum_{n=0}c_{n,\varphi}|n\rangle,
        \rho &=\sum_{n,n'}^{N_{\text{max}}} c_{n,n'} |n\rangle \langle n'|,
    \end{aligned}
\end{equation}
with $c_{n,\varphi}:=e^{in\varphi}H_n(x_\varphi)/\sqrt{2^nn!}$ and  $n$. Upon substituting Eq.~\ref{eq:fock_basis1} into \ref{eq: quad_distribution_1} we get:
\begin{equation}
    \begin{aligned}
        P(|x_\varphi|)\!&\approx\text{Tr}\!\left[\frac12(|x_\varphi\rangle \langle x_\varphi| + |-x_\varphi\rangle \langle- x_\varphi|) \sum_{n,n'}^{N_{\text{max}}} c_{n,n'}|n\rangle \langle n'|\right]\\
        &\approx\!\frac12\sum_{n,n'}^{N_{\text{max}}} c_{n,n'} (\langle n'|x_\varphi\rangle \langle x_\varphi|n\rangle + \langle n'|-x_\varphi\rangle \langle- x_\varphi|n\rangle)\\
        &\approx \! \frac12\sum_{n,n'}^{N_{\text{max}}} c_{n,n'} e^{-i(n'-n)\varphi} [H_{n}(x_\varphi)H_{n'}^*(x_\varphi)+H_n(-x_\varphi)H_{n'}^*(-x_\varphi)] \\
        &\approx \sum_{n,n'}^{N_{\text{max}}} c_{n,n'}e^{-i(n'-n)\varphi} \Pi^{n',n}(x_\varphi),
    \end{aligned}
\end{equation}
where we have defined $\Pi^{n',n}(x_\varphi):=H_{n}(x_\varphi)H_{n'}^*(x_\varphi)+H_n(-x_\varphi)H_{n'}^*(-x_\varphi)$, where $N_{\text{max}}$ represents the Fock-space truncation and where $c_{n,n'} = \langle n|\rho| n'\rangle$.

\section{Tomographic reconstruction with SDP}
\label{app:SDP}
We want to solve the following equation for the unknown coefficients $c_{n,n'}$:
\begin{equation}
    \label{eq:pb_ap}
    \begin{aligned}
P(|x_\varphi|)\!&\approx \sum_{n,n'}^{N_{\text{max}}} c_{n,n'}e^{-i(n'-n)\varphi} \Pi^{n',n}(x_\varphi),
    \end{aligned}
\end{equation}
where we have defined $\Pi^{n',n}(x_\varphi):=H_{n}(x_\varphi)H_{n'}^*(x_\varphi)+H_n(-x_\varphi)H_{n'}^*(-x_\varphi)$, where $N_{\text{max}}=N$ represents the Fock-space truncation and where $c_{n,n'} = \langle n|\rho| n'\rangle$.

The initial state $\rho$ is written in the photon number basis as:
\begin{equation}
\label{eq:fock_basis}
    \begin{aligned}
        \rho &=\sum_{n,n'}^{N_{\text{max}}} c_{n,n'} |n\rangle \langle n'|.
    \end{aligned}
\end{equation}
The sum $n, n'$ goes to infinity, but for practical purposes, we truncate it to a cutoff $N$ such that the terms for $(n, n') > N$ do not contribute significantly to the sum. Now the matrix $\rho \in \mathbb{C}^{(N) \times (N)}$ is rescaled to the Liouville operator $\bf{M}$ $\in \mathbb{C}^{(N^2) \times (1)}$, which is a column matrix with $N^2$ coefficients for the state $\rho$.
\\
Changing the pump phase $\varphi$, we obtain the quadrature distributions $P(x_{\varphi})\in \mathbb{R}^{(1) \times (N_{bins})}$ for each phase from experiments or simulations. Here, $N_{bins}$, $L$ ( Table~\ref{tbl:parameters}) represent the number of bins used in the histogram representation of $P(x_{\varphi})$ distributed in the range $(0, L)$ for OPA-based measurements and $(-L, L)$ for homodyne-based measurements. The collective matrix for quadrature distribution of $N_{phases}$ is a matrix with dimensions $(N_{phases}\times N_{bins})$. This is rescaled to the liouville vector representation $\bf{P}$$\in \mathbb{R}^{(N_{phases}N_{bins}) \times (1)}$.
\\

POVM elements $\Pi^{n',n}\in \mathbb{C}^{(N) \times (N)}(x_{\varphi})$ are calculated using hermite polynomials $H_n(x_\varphi)$ in an iterative process over $n, n'\in (N,N)$ for each phase given by:
\begin{equation}
    \begin{aligned}
            \Pi^{n',n}_{x_\varphi} =H_{n}(x_\varphi)H_{n'}^*(x_\varphi)+H_n(-x_\varphi)H_{n'}^*(-x_\varphi).
    \end{aligned}
\end{equation}
The distribution of $x_{\varphi}$ for a fixed $\phi$ is distributed in $N_{bins}$, and the value of $x_{\varphi}$ for each bin is the mean value of $x_{\varphi}$ in that bin. This gives us $\Pi^{n',n}(x_{\varphi})\in \mathbb{C}^{(N) \times (N)}$ for each bin. This is repeated for all phases ($N_{phases}$), which is then used to construct a multidimensional array with shape ($N_{phases}, N_{bins}, N, N$). This is further rescaled to a Liouville operator $\bf{\Pi}$ $\in \mathbb{C}^{(N_{phases}N_{bins}) \times (N^2)}$.

The problem in Eq.~\ref{eq:pb_ap} can now be written in the compact form $\bf{P}=\Pi M$, which in the matrix representation is written as: 

\begin{equation*}
    \begin{bmatrix}
        P^{1}(x_{\varphi_{1}})\\
        P^{2}(x_{\varphi_{1}})\\
        \vdots\\
        P^{N_{bins}}(x_{\varphi_{1}})\\
        \vdots\\
        P^{1}(x_{\varphi_{2}})\\
        \vdots\\
        P^{N_{bins}}(x_{\varphi_{2}})\\
        \vdots\\
        P^{N_{bins}}(x_{\varphi_{N_{phases}}})\\
        
    \end{bmatrix}
=
    \begin{bmatrix}
\Pi^{1,1}_{\varphi_{1}, 1} & \Pi^{1,2}_{\varphi_{1}, 1} & \dots & \Pi^{N,N}_{\varphi_{1}, 1} \\
\Pi^{1,1}_{\varphi_{1}, 2} & \Pi^{1,2}_{\varphi_{1}, 2} & \dots & \Pi^{N,N}_{\varphi_{1}, 2} \\
\vdots & \vdots & \vdots & \vdots \\
\Pi^{1,1}_{\varphi_{1}, N_{bins}} & \Pi^{1,2}_{\varphi_{1}, N_{bins}} & \dots & \Pi^{N,N}_{\varphi_{1}, N_{bins}} \\
\vdots & \vdots & \vdots & \vdots \\
\Pi^{1,1}_{\varphi_{N_{phases}}, 1} & \Pi^{1,2}_{\varphi_{N_{phases}}, 1} & \dots & \Pi^{N,N}_{\varphi_{N_{phases}}, 1} \\
\Pi^{1,1}_{\varphi_{N_{phases}}, 2} & \Pi^{1,2}_{\varphi_{N_{phases}}, 2} & \dots & \Pi^{N,N}_{\varphi_{N_{phases}}, 2} \\
\vdots & \vdots & \ddots & \vdots \\
\Pi^{1,1}_{\varphi_{N_{phases}}, N_{bins}} & \Pi^{1,2}_{\varphi_{N_{phases}}, N_{bins}} & \dots & \Pi^{N,N}_{\varphi_{N_{phases}}, N_{bins}} \\
    \end{bmatrix}
    \begin{bmatrix}
        c_{1,1}\\
        c_{1,2}\\
        c_{1,3}\\
        \vdots\\
        c_{1,N}\\
        c_{2,1}\\
        c_{2,2}\\
        \vdots\\
        c_{2,N}\\
        \vdots\\
        c_{N,N}
    \end{bmatrix},
    \end{equation*}
where $\bf{P}\in \mathbb{R}^{(N_{phases}N_{bins})\times1}$, $\bf{\Pi}\in \mathbb{C}^{(N_{phases}N_{bins})\times N^2}$ and $\bf{M}\in \mathbb{C}^{N^2\times1}$.

\section{Sign-free quadrature measurement estimates}
\label{app:direct}

Let $\rho$ be a quantum state and let $\eta\le1$. We denote by $(\varphi,x)\mapsto p_{\eta,\rho}(\varphi,x)$ the probability density function associated to the homodyne measurement with efficiency $\eta$ of the quadrature $\hat x_\varphi=\cos\varphi\,\hat q+\sin\varphi\,\hat p$ of the state $\rho=\sum_{m,n\ge0}\rho_{mn}\ket m\!\bra n$. We have~\cite{d2003quantum}:
\begin{equation}
    \begin{aligned}
        \rho_{n+d,n}&=\underset{\varphi,x}{\mathbb E}\left[\mathcal R^\eta_{n,n+d}\right]\\
        &=\int_0^\pi\frac{d\varphi}\pi\int_{-\infty}^{+\infty}dxp_{\eta,\rho}(\varphi,x)\mathcal \mathcal R^\eta_{n,n+d}(\varphi,x),
    \end{aligned}
\end{equation}
for all $n,d\in\mathbb N$, where
\begin{equation}\label{eq:defR}
    \begin{aligned}
        &\mathcal R^\eta_{n,n+d}(\varphi,x):=e^{id(\varphi+\frac\pi2)}\sqrt{\frac{n!}{(n+d)!}}\int_{-\infty}^{+\infty}dk|k|e^{-\left(1-\frac1{2\eta}\right)k^2-2ikx}k^dL_n^{(d)}(k^2),
    \end{aligned}
\end{equation}
where $L_n^{(d)}$ is a generalised Laguerre polynomial. As a result:
\begin{equation}
        \mathcal R^\eta_{n,n+d}(\varphi,-x)=(-1)^d\mathcal \mathcal R^\eta_{n,n+d}(\varphi,x).
\end{equation}
In particular, for $d=2k$,
\begin{equation}
    \mathcal R^\eta_{n,n+2k}(\varphi,-x)=\mathcal \mathcal R^\eta_{n,n+2k}(\varphi,x).
\end{equation}
We thus obtain
\begin{equation}\label{eq:expectedvparityapp}
    \rho_{n+2k,n}=\underset{\varphi,|x|}{\mathbb E}\left[\mathcal R^\eta_{n,n+2k}\right].
\end{equation}
%

\section{Similarity between elementwise and SDP reconstruction}
\begin{figure*}
    \centering
    \includegraphics[height=8cm, width = \textwidth]{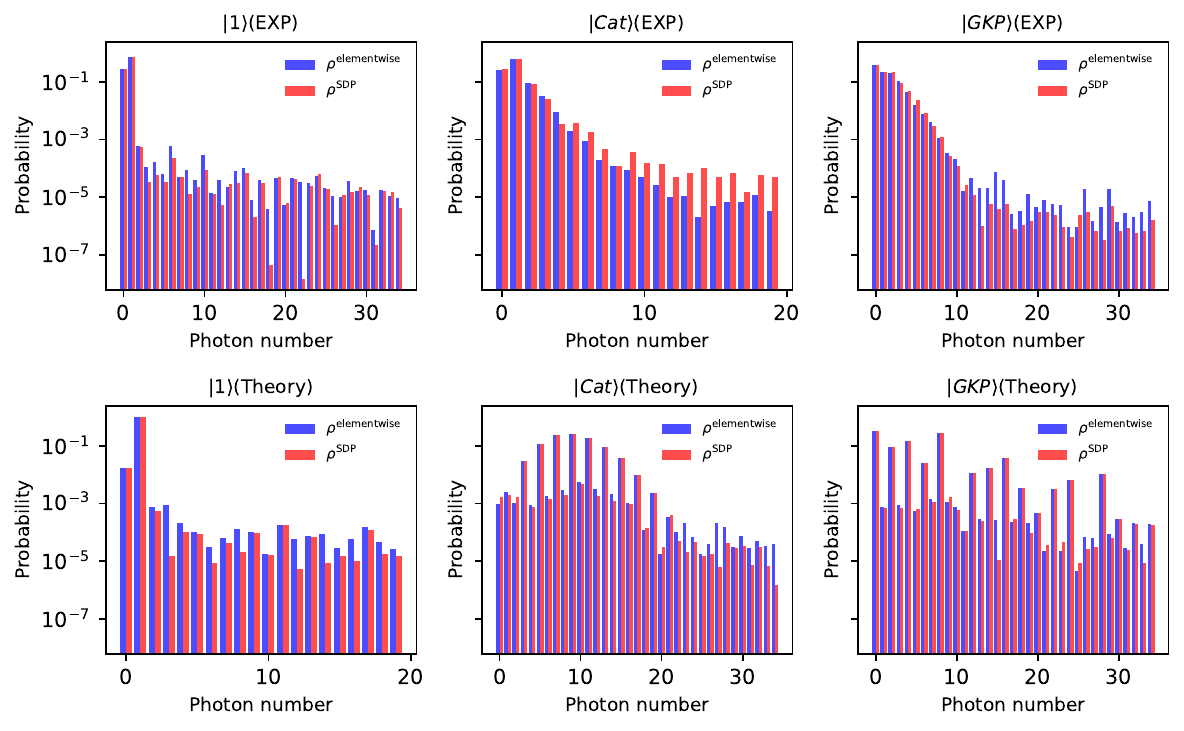}
    \caption{Diagonal elements of the reconstructed density matrix using elementwise reconstruction and SDP reconstruction. The top row shows the comparison for experimental states and the bottom row for the theoretically simulated single-photon state, cat state and the GKP state }
    \label{fig:rho_comparison}
\end{figure*}

Here, we show the similarity between the reconstructed states using the SDP method with the ideal elementwise reconstructed single-photon, small amplitude cat state and GKP state. In Fig.\ \ref{fig:rho_comparison} we plot the diagonal elements of the reconstructed states for experimental states in the top row and theoretically simulated states in the bottom row. We can see that the probabilities match even for appreciable fock space basis overlap.
\section{Stellar rank}
\label{app:stellar}

The stellar rank $r^\star$ is a non-Gaussian measure of quantum states, which for single-mode pure states is defined as the number of zeros of their Husimi $Q$-function~\cite{chabaud2020stellar} (with half-multiplicity). Importantly, a single-mode quantum state $\ket\psi$ of stellar rank $r^\star(\psi)=k$ can be decomposed as
\begin{equation}
    \ket\psi=\hat D(\alpha)\hat S(\xi)\sum_{n=0}^kc_n\ket n,
\end{equation}
where $\alpha,\xi\in\mathbb C$ and where $\ket{C_\psi}:=\sum_{n=0}^kc_n\ket n$ is a normalized state of finite support over the Fock basis (a so-called core state). $\hat S(\xi)$ is the squeezing operator. The stellar rank is invariant under Gaussian unitaries, non-increasing under Gaussian channels~\cite{chabaud2021holomorphic}, and corresponds to the minimum number of elementary non-Gaussian operations (photon-addition or photon-subtraction) necessary to engineer a state from the vacuum, together with Gaussian unitary operations. Any quantum state may be approximated to arbitrary precision by states of finite stellar ranks. Moreover, for any fixed stellar rank, the optimal approximation can be obtained by optimization over two complex parameters~\cite[Theorem~1]{chabaud2021certification}: the maximum achievable fidelity with a target pure state $\ket\psi$ using states of rank less than or equal to $k$ is given by
\begin{equation}
    \sup_{r^\star(\rho)\le k}F(\rho,\psi)=\sup_{\hat G\in\mathcal G}\Tr\left(\Pi_k\hat G\ket\psi\!\bra\psi\hat G^\dag\right),
\end{equation}
where $\Pi_k=\sum_{n\le k}\ket n\!\bra n$ and where the supremum is over single-mode Gaussian unitary operations, which can be parametrized by two complex parameters. Moreover, assuming the optimization yields a Gaussian operation $\hat G_0$, an optimal approximation of stellar rank $k$ of the state $\ket\psi$ is given by
\begin{equation}
    \frac{\hat G_0^\dag\Pi_k\hat G_0\ket\psi}{\|\Pi_k\hat G_0\ket\psi\|}.
\end{equation}

This result, which can be extended to multimode states~\cite{chabaud2021certification}, allows us to obtain the profile of stellar fidelities of any target pure state, which is defined as the list of the maximum achievable fidelities with that target state using states of finite stellar ranks \cite{hahn2024assessing} (see \cite{codestellarprofile} for the numerical implementation, as well as \cite{nauth2026optimal,wang2026bosonic}). These profiles provide a qualitative description of the difficulty to prepare a quantum state in terms of elementary non-Gaussian operations, such as photon-addition or photon-subtraction. Furthermore, the stellar robustness profiles may be used to witness the stellar rank of an experimental state by direct fidelity estimation. We give various examples in the following section.


\section{Upper bound on truncated fidelity}
\label{app:Fidelity_bd}
First, we explicitly write the even cat state as an infinite sum in the Fock basis. We then define the cutoff version of this state by terminating the sum at a cutoff $N_0$ and renormalizing:

\begin{equation}
|\text{cat}_{\alpha}^{+}\rangle = \frac{1}{\sqrt{\cosh(|\alpha|^{2})}}\sum_{n=0}^{+\infty}\frac{\alpha^{2n}}{\sqrt{(2n)!}}|2n\rangle
\end{equation}

\begin{equation}
|\text{cat}_{\alpha,N_{0}}^{+}\rangle = \frac{1}{\sqrt{\mathcal N(N_0)}}\sum_{n=0}^{N_{0}}\frac{\alpha^{2n}}{\sqrt{(2n)!}}|2n\rangle
\end{equation}
where $\mathcal N(N_0)=\sum_{n=0}^{N_{0}}\frac{|\alpha|^{4n}}{(2n)!}$ is a normalization factor and with
\begin{equation}
    \cosh(|\alpha|^2) = \sum_{n=0}^{\infty} \frac{|\alpha|^{4n}}{(2n)!}
\end{equation}

The fidelity $F$ between the ideal state and the truncated state is the square of their inner product. We have
\begin{equation}
\langle \text{cat}_{\alpha,N_0}^{+} | \text{cat}_{\alpha}^{+} \rangle = \frac{1}{\sqrt{\cosh(|\alpha|^{2})\mathcal N(N_0)}} \sum_{n=0}^{N_0} \frac{|\alpha|^{4n}}{(2n)!}=\sqrt{\frac{\sum_{n=0}^{N_0} \frac{|\alpha|^{4n}}{(2n)!}}{\cosh(|\alpha|^{2})}},
\end{equation}
so the fidelity is
\begin{equation}
F = \frac{\sum_{n=0}^{N_0} \frac{|\alpha|^{4n}}{(2n)!}}{\cosh(|\alpha|^{2})}.
\end{equation}

Using the pure state relation $D^2 = 1 - F$:
\begin{equation}
D^2 = 1 - \frac{\sum_{n=0}^{N_0} \frac{|\alpha|^{4n}}{(2n)!}}{\cosh(|\alpha|^2)} = \frac{\cosh(|\alpha|^2) - \sum_{n=0}^{N_0} \frac{|\alpha|^{4n}}{(2n)!}}{\cosh(|\alpha|^2)}.
\end{equation}
Since $\cosh(|\alpha|^2)\ge1$ we obtain:
\begin{equation}
D^2 \le \sum_{n=N_0+1}^{\infty} \frac{|\alpha|^{4n}}{(2n)!}
\end{equation}

Finally, we apply the Lagrange form of the Taylor remainder for the function $f(x) = \cosh(x)$ at $x = |\alpha|^2$: for a truncation at degree $2N_0$, the remainder $R$ is:

\begin{equation}
R_{2N_0}(|\alpha|^2) = \frac{f^{(2N_0+1)}(\xi)}{(2N_0+1)!} (|\alpha|^2)^{2N_0+1},
\end{equation}
for some value $\xi \in (0, |\alpha|^2)$. The $(2N_0+1)$-th derivative of $\cosh$ is $\sinh$. We bound $\sinh(\xi)$ with $e^{|\alpha|^2}$, yielding the final result:

\begin{equation}
D(|\text{cat}_{\alpha,N_{0}}^{+}\rangle,|\text{cat}_{\alpha}^{+}\rangle) \le \sqrt{\frac{|\alpha|^{4N_{0}+2}e^{|\alpha|^{2}}}{(2N_{0}+1)!}}.
\end{equation}

\section{Profiles of stellar fidelities}
\label{app:profiles}

In this section, we derive the profiles of stellar fidelities for binomial code states, cat states, and GKP states. A library for computing these profiles numerically can be found here \cite{codestellarprofile}.

%
\begin{figure}[t]
	\begin{center}
		\includegraphics[width=0.7\columnwidth]{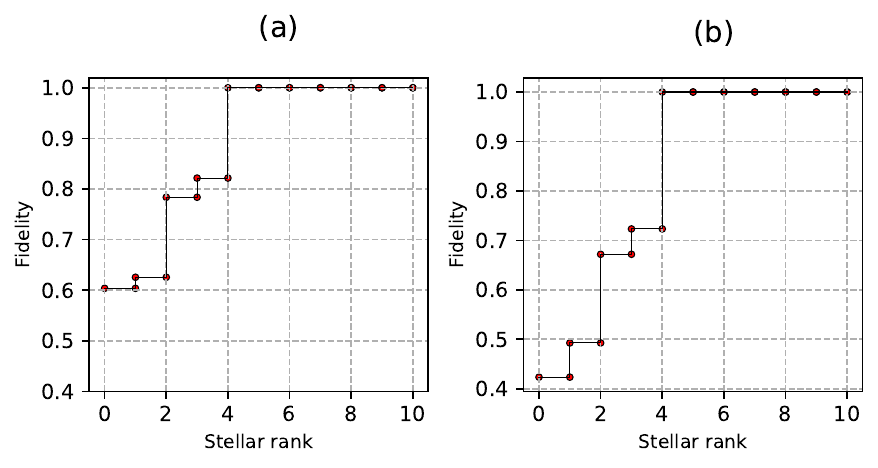}
		\caption{Achievable fidelities for a target binomial state $\ket{W_{\text{even}}(1,1)}$, $\ket{W_{\text{even}}(2,1)}$ respectively. For each rank $k\in\mathbb N$ on the horizontal axis, the vertical line depicts the achievable fidelities between states of stellar rank less or equal to $k$ and the target state.}
		\label{fig:Rbinom}
	\end{center}
\end{figure}

\subsection{Binomial States}
\label{app:binom}

Binomial states are defined as~\cite{PhysRevX.6.031006}:
\begin{equation}
    \ket{W_{\text{even/odd}}(N,S)}=\frac1{\sqrt{2^N}}\!\!\!\!\sum_{\substack{p=0\\p\text{ even/odd}}}^{N+1}\!\!\!\!\sqrt{\binom{N+1}p}\ket{p(S+1)}\!,
\end{equation}
where $N,S\in\mathbb N$. We have computed numerically the profile of achievable fidelities with the binomial state $\ket{W_{\text{even}}(2,1)}=\frac12\ket0+\frac{\sqrt3}2\ket4$. This profile is depicted in Fig.~\ref{fig:Rbinom}. Note how the maximum achievable fidelity increases more from odd to even ranks than from even to odd ranks. This is due to the target state having support only on even Fock states.


\subsection{Cat states}
\label{app:cat}

Cat states of amplitude $\alpha$ are defined as:
\begin{equation}
\ket{\text{cat}^\pm_\alpha}=\frac1{\sqrt{\mathcal N^\pm_\alpha}}(\ket\alpha\pm\ket{-\alpha}),
\end{equation}
where $\mathcal N^\pm_\alpha=2(1\pm e^{-2|\alpha|^2})$ is a normalising constant. These states have an infinite stellar rank and thus cannot be perfectly approximated using states of bounded stellar rank.

\begin{figure*}
	\begin{center}
		\includegraphics[width = 0.7\columnwidth]{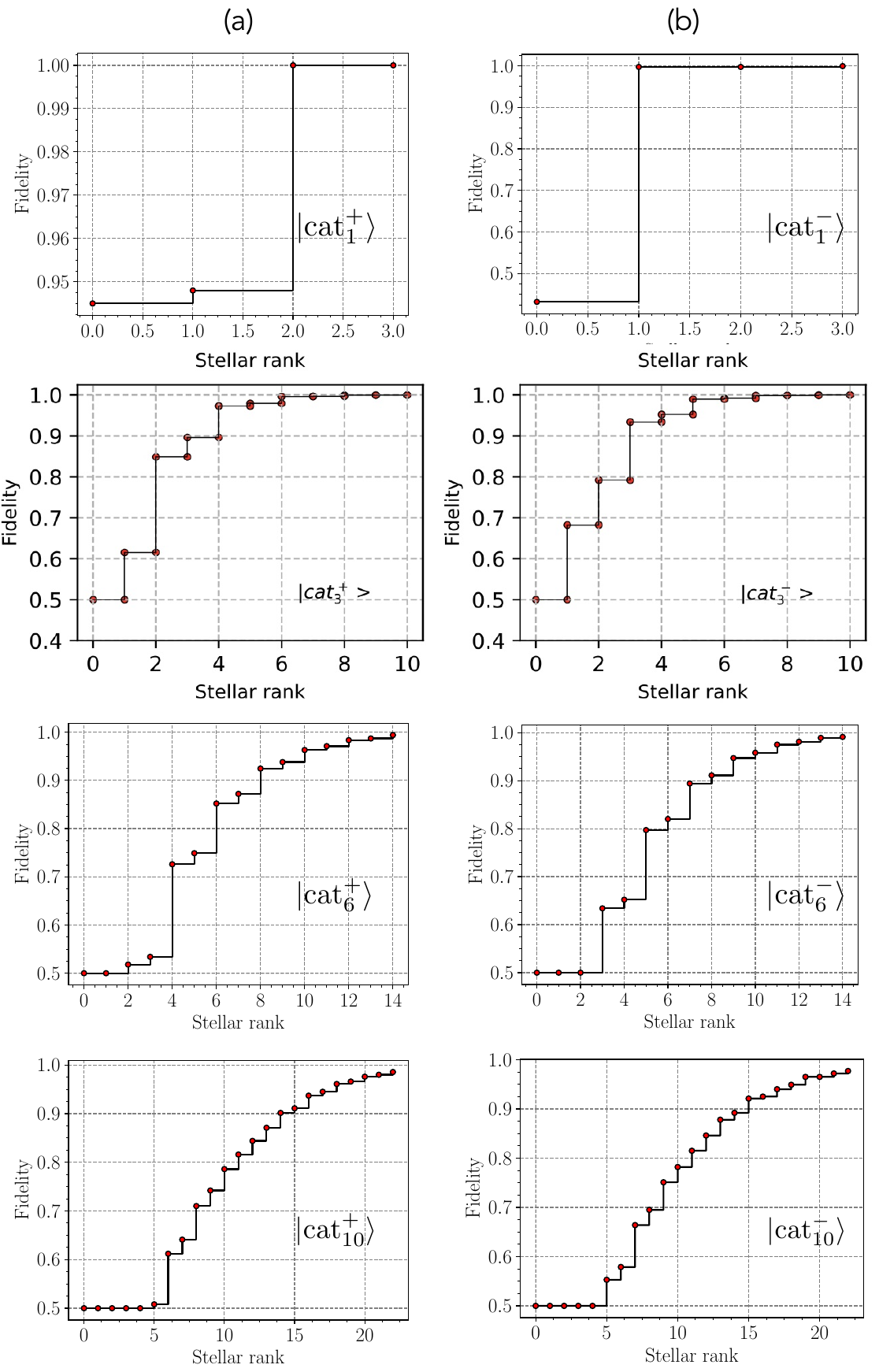}
		\caption{Achievable fidelities for (a) target cat$^+$ states and (b) cat$^-$ states, of amplitudes $1$, $3$, $6$ and $10$. For each rank $k\in\mathbb N$ on the horizontal axis, the vertical line depicts the achievable fidelities between states of stellar rank less or equal to $k$ and the target state.}
		\label{fig:Rcat}
	\end{center}
\end{figure*}
\begin{figure*}
	\begin{center}
		\includegraphics[width = 0.7\columnwidth]{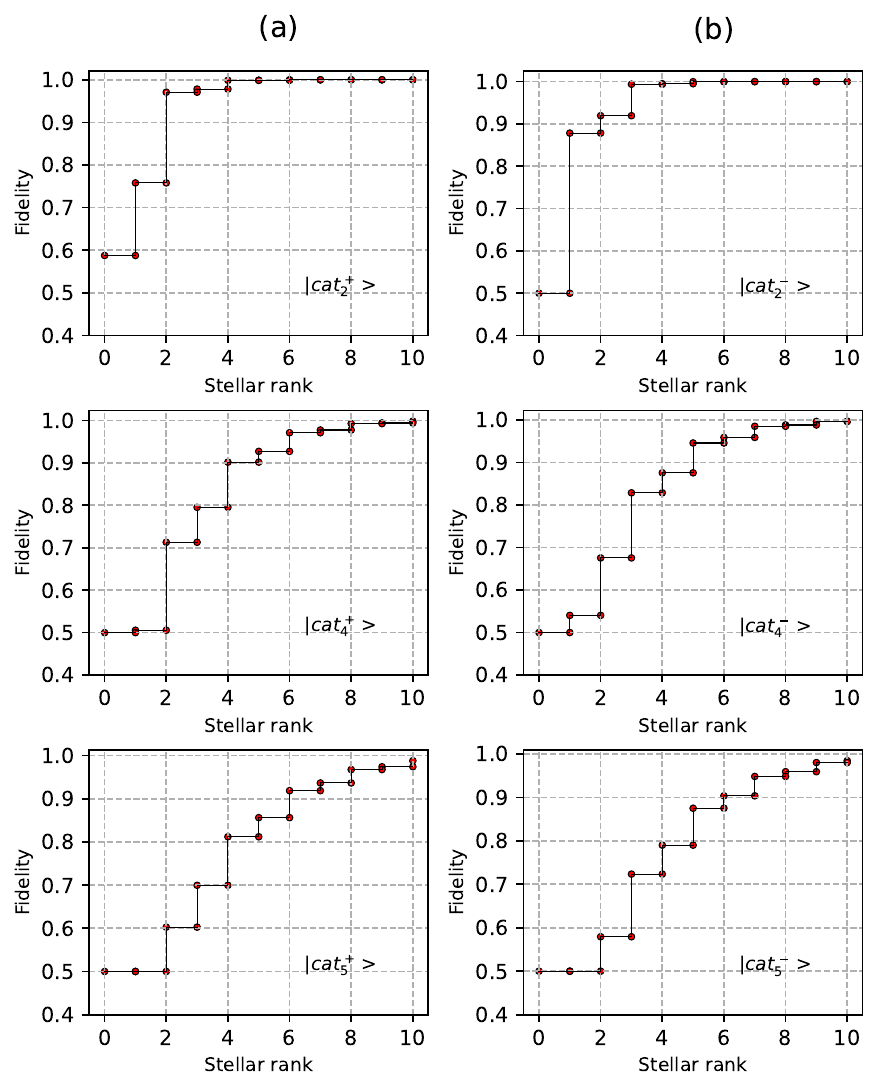}
		\caption{Achievable fidelities for (a) target cat$^+$ states and (b) cat$^-$ states, of amplitudes $2$, $4$, and $5$. For each rank $k\in\mathbb N$ on the horizontal axis, the vertical line depicts the achievable fidelities between states of stellar rank less or equal to $k$ and the target state.}
		\label{fig:Rcat2}
	\end{center}
\end{figure*}
We have computed numerically the profile of achievable fidelities with cat states of several amplitudes. These profiles are depicted in Fig.~\ref{fig:Rcat} for cat$^+$ and cat$^-$ states. From the numerics and the obtained profiles, we make various observations:\\

\begin{enumerate}

\item The main difference between low amplitude cat$^+$ and cat$^-$ states is that the former are easier to approximate by low stellar rank states than the latter. This is due to low amplitudes cat$^+$ states being close to the vacuum state and low amplitudes cat$^-$ states being close to a single-photon Fock state.

\item High amplitude cat states are `more non-Gaussian' than low amplitude cat states, in the sense that more photon additions/subtractions are needed to approximate them to the same precision.

\item The maximum achievable fidelity increases more from odd to even ranks (resp.\ even to odd ranks) than from even to odd ranks (resp.\ odd to even ranks) for cat$^+$ states (resp.\ cat$^-$ states). Like for binomial states, this is due to cat$^+$ states (resp.\ cat$^-$ states) having support only on even (resp.\ odd) Fock states.

\item For each given amplitude, there is a critical stellar rank after which good approximation of the cat state becomes possible, which outlines the hardness of engineering cat states of high amplitude in terms of non-Gaussian elementary optical operations, namely photon-addition/subtractions. Before that critical stellar rank, the best Gaussian operation in the optimisation is either a displacement of the amplitude of the cat, or a squeezing, depending on the parity of the rank. Past that critical stellar rank, it is a smaller displacement combined with a squeezing.

\end{enumerate}

\twocolumngrid
\begin{figure}[htbp]
    \centering
    \includegraphics[width=\columnwidth]{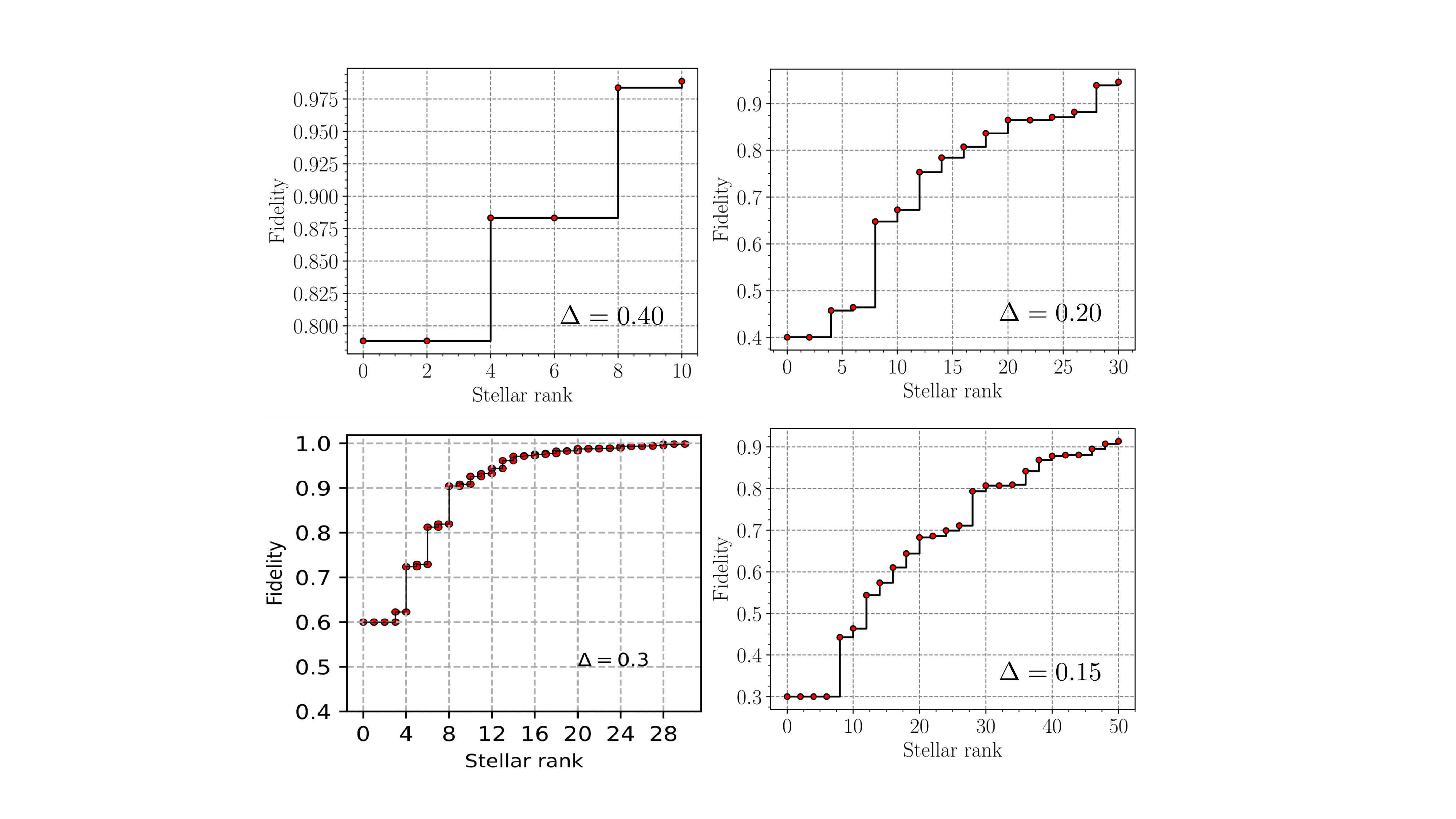}
    \caption{Achievable fidelities for target states $\ket{\text{GKP}}_\Delta$ with $j=0$, $d=2$ and $\Delta=\kappa$, for different values of $\Delta$ ($0.40$, $0.30$, $0.20$, $0.15$). For each rank $k\in\mathbb N$ on the horizontal axis, the vertical line depicts the achievable fidelities between states of stellar rank less or equal to $k$ and the target state. For clarity, only even ranks are depicted since we observe numerically that the highest achievable fidelities for odd stellar ranks are close to those of the previous even stellar rank.}
    \label{fig:RGKP}
\end{figure}
\onecolumngrid
\subsection{GKP states}


\subsubsection{Ideal GKP states}

The ideal (square lattice) GKP states are defined as:
\begin{equation}
    \ket{\text{GKP}_j}\propto\sum_{s\in\mathbb Z}\left|\sqrt{\frac{2\pi}d}(ds+j)\right\rangle_{\!\hat q},
\end{equation}
where $j\in\{0,\dots,d-1\}$ and $d\ge2$ is the dimension of the logical Hilbert space. These states have infinite norm. Their stellar function is given by \cite{chabaud2021continuous}:
\begin{equation}
    \begin{aligned}
        F^\star_{\text{GKP}_j}(z)&\propto\sum_{s\in\mathbb Z}e^{-\frac12z^2+2\sqrt{\frac\pi d}(ds+j)z-\frac\pi d(ds+j)^2}\\
        &=e^{-\frac12z^2+2\sqrt{\frac\pi d}jz-\frac\pi dj^2}\sum_{s\in\mathbb Z}e^{(2z\sqrt{\pi d}-2\pi j)s-\pi ds^2}\\
        &=e^{-\frac12z^2+2\sqrt{\frac\pi d}jz-\frac\pi dj^2}\vartheta(2z\sqrt{\pi d}-2\pi j,-\pi d),
    \end{aligned}
\end{equation}
where $\vartheta(z,\tau):=\sum_{s\in\mathbb Z}e^{zs}e^{\tau s^2}$ is the Jacobi theta function.


\subsubsection{Approximate GKP states}

We write $\hat q=\frac1{\sqrt2}(\hat a+\hat a^\dag)$ and $\hat p=\frac1{i\sqrt2}(\hat a-\hat a^\dag)$ with the convention $\hbar=1$, such that $[\hat a,\hat a^\dag]=\mathbb I$.
Let $\hat D(\alpha):=e^{\alpha\hat a^\dag-\alpha^*\hat a}$, $\hat S(\xi):=e^{\frac12(\xi\hat a^2-\xi^*\hat a^{\dag2})}$ and $\hat R(\varphi):=e^{i\varphi\hat a^\dag\hat a}$. Approximate GKP states are obtained by replacing the position eigenstates in the infinite sum by finitely squeezed coherent states with a Gaussian weight:
\begin{equation}\label{eq:approxGKP}
        \ket{\text{GKP}_{j,\Delta,\kappa}}:=\frac1{\sqrt{N_{j,\Delta,\kappa}}}\sum_{s\in\mathbb Z}{e^{-\frac{\pi\kappa^2}d(ds+j)^2}}\hat D\left(\!\sqrt{\frac\pi d}(ds+j)\!\right)\!\hat S(-\ln\Delta)\ket0_{vac}\!.
\end{equation}
A squeezed coherent state $\hat D(\alpha)\hat S(-\ln\Delta)\ket0_{vac}$ has the position representation:
\begin{equation}
    \prescript{}{\hat q}{\bra q}\hat D(\alpha)\hat S(-\ln\Delta)\ket0_{vac}=\left(\pi\Delta^2\right)^{-1/4}e^{-{\frac{(q-q_0)^2}{2\Delta^2}}+ip_0q},
\end{equation}
where $\alpha:=\frac1{\sqrt2}(q_0+ip_0)$.
Hence, the normalisation $N_{j,\Delta,\kappa}$ in Eq.~\ref{eq:approxGKP} is given by:
\begin{widetext}
\begin{equation}
    \begin{aligned}
        N_{j,\Delta,\kappa}&=\sum_{s,t\in\mathbb Z}e^{-\frac{\pi\kappa^2}d[(ds+j)^2+(dt+j)^2]}\prescript{}{vac}{\bra0}\hat S(\ln\Delta)\hat D\left(\sqrt{\frac\pi d}(ds-dt)\right)\hat S(-\ln\Delta)\ket0_{vac}\\
        &=\sum_{s,t\in\mathbb Z}e^{-\frac{\pi\kappa^2}d[(ds+j)^2+(dt+j)^2]}\left(\pi\Delta^2\right)^{-1/2}\int_{q\in\mathbb R}e^{-{\frac{q^2}{2\Delta^2}}}e^{-{\frac{(q-\sqrt{2\pi d}(s-t))^2}{2\Delta^2}}}dq\\
        &=\sum_{s,t\in\mathbb Z}e^{-\frac{\pi\kappa^2}d[(ds+j)^2+(dt+j)^2]}e^{-\frac{\pi d}{2\Delta^2}(s-t)^2}\left(\pi\Delta^2\right)^{-1/2}\int_{q\in\mathbb R}e^{-{\frac{(q-\sqrt{\frac{\pi d}2}(s-t))^2}{\Delta^2}}}dq\\
        &=\sum_{s,t\in\mathbb Z}e^{-\frac{\pi\kappa^2}d[(ds+j)^2+(dt+j)^2]}e^{-\frac{\pi d}{2\Delta^2}(s-t)^2}.
    \end{aligned}
\end{equation}
\end{widetext}

Different notions of approximate GKP states have been introduced, three of which have been proven equivalent, up to a squeezing operation~\cite{matsuura2020equivalence}.

We now turn to the question underlying the stellar robustness profile of GKP states: for each $N\in\mathbb N^*$, what is the best approximation of stellar rank lower than $N$ of an approximate GKP state? Equivalently, what is the state of the form $\hat S(\xi)\hat D(\alpha)\sum_{n=0}^{N-1}c_n\ket n$ that achieves the highest fidelity with the state $\ket{\text{GKP}_{j,\Delta,\kappa}}$? 

In~\cite{tzitrin2020progress}, lower bounds on the optimal fidelities were obtained by assuming $\xi\in\mathbb R$, $\alpha=0$ and $c_n=0$ for $n$ odd, by running a global optimization over the remaining parameters (thus scaling with $N$).
As it turns out, the optimal parameters $\xi$, $\alpha$ and $(c_n)_{0\le n<N}$ can be obtained by optimizing only over the complex Gaussian parameters $\xi$ and $\alpha$, using Theorem 1 of~\cite{chabaud2021certification} (see Appendix~\ref{app:stellar}). The optimal fidelity is then given by:
\begin{equation}\label{eq:overlaps}
    \begin{aligned}
    &\sup_{r^\star(\rho)<N}F(\rho,\ket{\text{GKP}_{j,\Delta,\kappa}})\\&\quad=\sup_{\zeta,\alpha\in\mathbb C}\sum_{n=0}^{N-1}\left|\braket{n|\hat D(\alpha)\hat S(\zeta)|\text{GKP}_{j,\Delta,\kappa}}\right|^2\\
    &\quad=\sup_{\xi,\alpha\in\mathbb C}\sum_{n=0}^{N-1}\left|\braket{n|\hat D(\alpha)\hat S(\xi)\hat S(\ln\Delta)|\text{GKP}_{j,\Delta,\kappa}}\right|^2,
    \end{aligned}
\end{equation}
Where in the last line, we used the fact that the supremum is over all Gaussian unitaries in order to obtain more convenient expressions later.
This can, in turn, be expressed using
\begin{equation}\label{eq:nderiv}
    \braket{n|\psi}=\frac1{\sqrt{n!}}\partial_z^nF^\star_\psi(z)|_{z=0},
\end{equation}
for $\ket\psi=\hat D(\alpha)\hat S(\xi)\hat S(\ln\Delta)\ket{\text{GKP}_{j,\Delta,\kappa}}$. 

We use the notation $t_\xi:=-e^{-i\theta}\tanh r$, and we look for $\chi_s,\gamma_s\in\mathbb C$ such that
\begin{equation}
    \begin{aligned}
        &\hat D(\alpha)\hat S(\xi)\hat S(\ln\Delta)\hat D\left(\sqrt{\frac\pi d}(ds+j)\right)\hat S(-\ln\Delta)\ket0_{vac}\\
        &\quad\quad=e^{\chi_s}\hat D(\gamma_s)\hat S(\xi)\ket0_{vac}.
    \end{aligned}
\end{equation}
To that end, we make use of the relations~\cite{leonhardt2010essential}:
\begin{equation}
    \begin{aligned}
        \hat S(\mu)\hat D(\beta)\hat S(-\mu)&=\hat D[\cosh|\mu|(\beta+\beta^*t_\mu)]\\
        \hat D(\alpha)\hat D(\beta')&=e^{\frac12(\alpha{\beta'}^*-\alpha^*\beta')}\hat D(\alpha+\beta')
    \end{aligned}
\end{equation}
With $\mu=\ln\Delta=(-\ln\Delta)e^{i\pi}\in\mathbb R$ and setting $\beta_s:=\sqrt{\frac\pi d}(ds+j)\in\mathbb R$ we obtain
\begin{equation}\label{eq:chigamma}
    \begin{aligned}
        \chi_s&=\frac{\beta_s}{2\Delta}\cosh|\xi|[\alpha(1+t_\xi^*)-\alpha^*(1+t_\xi)]\\
        \gamma_s&=\alpha+\frac{\beta_s}\Delta\cosh|\xi|(1+t_\xi).
    \end{aligned}
\end{equation}
In particular, with Eq.~\ref{eq:approxGKP}, we obtain
\begin{equation}\label{eq:superposG}
    \begin{aligned}
        &\hat D(\alpha)\hat S(\xi)\hat S(\ln\Delta)\ket{\text{GKP}_{j,\Delta,\kappa}}\\
        &=\frac1{\sqrt{N_{j,\Delta,\kappa}}}\sum_{s\in\mathbb Z}{e^{-\frac{\pi\kappa^2}d(ds+j)^2}}e^{\chi_s}\hat D(\gamma_s)\hat S(\xi)\ket0_{vac}.
    \end{aligned}
\end{equation}
The stellar function is linear with respect to superpositions, and the stellar function of a Gaussian state $\hat D(\gamma)\hat S(\xi)\ket0_{vac}$ is given by~\cite{chabaud2020stellar}:
\begin{equation}
    \begin{aligned}
        F^\star_{\gamma,\xi}(z)&=(1-|t_\xi|^2)^{1/4}e^{\frac12t_\xi z^2+(\gamma-t_\xi\gamma^*)(z-\frac12\gamma^*)}\\
        &=\frac1{\sqrt{\cosh|\xi|}}e^{\frac12t_\xi \left[z-(\gamma^*-\frac\gamma{t_\xi})\right]^2+\frac12\gamma(\gamma^*-\frac\gamma{t_\xi})}.
    \end{aligned}
\end{equation}
With Eq.~\ref{eq:nderiv}, we have:
\begin{equation}\label{eq:Hermite_nDS}
    \begin{aligned}
        &\braket{n|\hat D(\gamma)\hat S(\xi)|0}_{vac}\\        &=\frac1{\sqrt{n!}}\partial_z^nF^\star_{\gamma,\xi}(z)|_{z=0}\\
        &=\frac1{\sqrt{n!\cosh|\xi|}}e^{\frac12\gamma(\gamma^*-\frac\gamma{t_\xi})}\partial_z^n\left[e^{\frac12t_\xi\left[z-(\gamma^*-\frac\gamma{t_\xi})\right]^2}\right]_{z=0}\\
        &=\sqrt{\frac{(-t_\xi)^n}{2^nn!\cosh|\xi|}}e^{\frac12t_\xi\gamma^*(\gamma^*-\frac\gamma{t_\xi})}H_n\left(\sqrt{-\frac12t_\xi}(\gamma^*-\frac\gamma{t_\xi})\right),
    \end{aligned}
\end{equation}
where we used $\partial_z^n[e^{-a(z-b)^2}]_{z=0}=e^{-ab^2}a^{n/2}H_n(\sqrt ab)$, with $H_n$ the $n^{th}$ Hermite polynomial.

Combining Eqs.~\ref{eq:superposG} and~\ref{eq:Hermite_nDS} yields developed expressions for the overlaps in Eq.~\ref{eq:overlaps}.
Using these expressions, we have obtained the robustness profile of GKP state depicted in Fig.~\ref{fig:RGKP}.

\section{Xanadu's GKP State Reconstruction}
\label{app:xanadu}
\begin{figure*}[!htb]
    \centering
    \includegraphics[width = 0.9\textwidth]{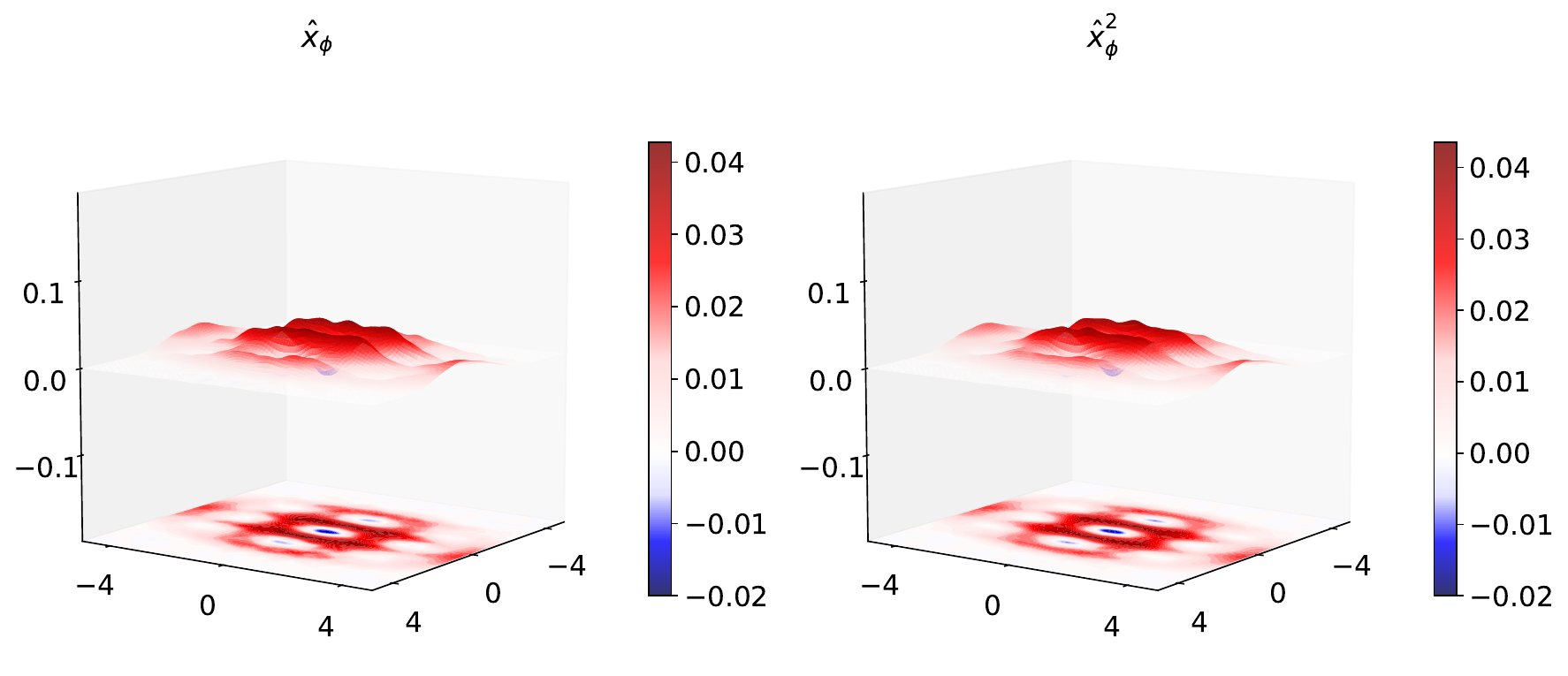}
    \caption{SDP reconstruction of the Wigner function using (left) complete quadrature data, and (right) sign-free quadrature data for Xanadu's experimental GKP state \cite{Larsen2025}.} 
    \label{fig:sim_states_xanadu}
\end{figure*}
Figure~\ref{fig:sim_states_xanadu} shows the reconstructed Wigner function of Xanadu's GKP state~\cite{Larsen2025}, obtained from experimental quadrature data using our SDP-based reconstruction method. We apply the technique to both the complete quadrature dataset, SDP($\hat{x}_{\phi}$), and the sign-free dataset, SDP($\hat{x}^2_{\phi}$). In both cases, the reconstructed states achieve a fidelity exceeding $99\%$ with the experimentally reported state.

To further characterize the state, we estimate its stellar rank by comparing it to the pure target state the experiment aimed to produce. To this end, we simulate the corresponding noiseless (loss-free) state using the circuit parameters reported in Ref.~\cite{Larsen2025} and compute its fidelity with the reconstructed experimental state. The resulting fidelity is comparatively low, reflecting the substantial optical losses present in the measurement setup. Reducing these losses would bring the experimental state closer to the intended pure target, revealing stronger non-Gaussian features and a correspondingly higher stellar rank.
\end{document}